\documentclass[12pt]{article}
\usepackage{amsmath}
\usepackage{graphicx}
\usepackage{natbib}
\usepackage{url} 
\usepackage{amssymb}
\usepackage{amsfonts}
\usepackage{pdfpages}

\usepackage{amssymb,amsmath,amsfonts,bbm}
\usepackage{mathrsfs}
\usepackage{calligra}
\usepackage{multirow}
\usepackage{amsthm}

\newcommand{\blind}{0}
\newtheorem{pro}{Proposition}
\theoremstyle{definition}
\newtheorem{rem}{Remark}
\newtheorem{as}{Assumption}
\newcommand{\PMB}[1]{% Poor Man's Bold
\leavevmode
\setbox0=\hbox{#1}%
\kern-0.02em\copy0\kern-\wd0
\kern0.02em\copy0\kern-\wd0
\kern-0.025em\raise0.0167em\box0
\kern-0.025em\raise0.0433em\box0}

\begin{document}

\def\spacingset#1{\renewcommand{\baselinestretch}%
{#1}\small\normalsize} \spacingset{1}

%%%%%%%%%%%%%%%%%%%%%%%%%%%%%%%%%%%%%%%%%%%%%%%%%%%%%%%%%%%%%%%%%%%%%%%%%%%%%%

\if0\blind
{
  \title{\bf Validation of Association}
  \author{Bogdan \'Cmiel\\
    Faculty of Applied Mathematics, AGH University of Science and Technology,\\
		Al. Mickiewicza 30, 30-059 Cracov, Poland\\
		{\it e-mail}: {\tt cmielbog@gmail.com}\\
    and \\
    Teresa Ledwina \\
    Institute of Mathematics, Polish Academy of Sciences\\
		ul. Kopernika 18, 51-617 Wroc{\l}aw, Poland\\
    {\it e-mail}: {\tt ledwina@impan.pl}}
  \maketitle
} \fi

\if1\blind
{
  \bigskip
  \bigskip
  \bigskip
  \begin{center}
    {\LARGE\bf Title}
\end{center}
  \medskip
} \fi

\bigskip
\begin{abstract}
Recognizing, quantifying and visualizing associations between two variables is increasingly important. This paper investigates how a new function-valued measure of dependence, the quantile dependence function, can be used to construct tests for independence and to provide an easily interpretable diagnostic plot of existing departures from the null model. %Three new tests are proposed and their consistency is studied.
The dependence function is designed to detect general dependence structure between variables in quantiles of the joint distribution. It gives an insight into how the dependence structures changes in different parts of the joint distribution. We define new estimators of the dependence function, discuss some of their properties, and apply them to construct new tests of independence. 
Numerical evidence is given on the test's benefits against three recognized independence tests introduced in the previous years. In real-data analysis, we illustrate the use of our tests and the graphical presentation of the underlying dependence structure.
\end{abstract}

\noindent%
{\it Keywords:}  Copula; Cross-quantilogram; Independence testing; Measure of dependence; Quantile,  Weighted statistics
\vfill

\newpage
\spacingset{1.5} % DON'T change the spacing!
\section{Introduction}
\label{S1}
\vspace{-0.2cm}
Measuring dependence and testing for independence have been intensively studied over the recent years. On the one hand, in many applied sciences it is a fundamental question how to quantify the dependence between variables under study. On the other hand, the present-day knowledge demonstrates that classical solutions are designed to capture specific, relatively simple, structures of dependence between two random variables. Thus they are not well suited to the scope of modern statistical analysis, and therefore may lead to completely misleading conclusions.
Another inspiration is exploratory data analysis, where one of the goals is to investigate a large data set and search for pairs of variables which are closely associated. Obviously, the dependence structure of random vector cannot be neglected in reliable data analysis. In particular, this problem is crucial in insurance and finance. For example, \cite{r58} showed that substantial deviations in the fair price of stop-loss premiums may occur even when there  are small departures from independence. For further evidence and related discussion see \cite{r61}, \cite{r59}, and \cite{r60}.

These challenges have stimulated great development in the area of evaluation of relationships, measuring their strength and testing for a lack of dependence via pertinent statistics. An emphasis has been put on capturing complex dependence structures which cannot be detected by classical solutions. 
For an illustration of several ideas and approaches to quantifying dependence and detecting associations which have been considered over the last decade (in chronological order), see: \cite{r46}, \cite{r8}, \cite{r35}, \cite{r32}, \cite{r54}, \cite{r18}, \cite{r25}, \cite{r34}, %Seidinovic et al. (2013), 
\cite{r2}, \cite{r49}, \cite{r27}, \cite{r19}, \cite{r36}, \cite{r1}, \cite{r11}, \cite{r55}, \cite{r50}, \cite{r52}, \cite{r37}, \cite{r51}, \cite{r53}. These papers also discuss many earlier ideas and developments. Most of them concern bivariate vectors and independent observations. A parallel stream of articles deals with dependent data and multivariate observations. It is beyond the scope of this paper to survey these more general situations. 

The present paper proposes and studies new tests of independence which are closely related to the function-valued measure of dependence $q$ proposed in \cite{r26,r27}.
The measure is conceptually appealing and has a straightforward interpretation facilitating a comprehensive view of the association structure.
 The pertaining tests are of simple form and have good power. Their extension to multivariate observations, dependent data, and for detecting positive or negative dependence is straightforward. Additionally, their big advantage is a natural link to an easily interpretable diagnostic plot. Consequently, in the case of rejection of the hypothesis on independence, reliable  information on which part of the population --- and to which extent --- invalidates the hypothesis is available.  

In many practical situations variables are measured in different scales. Therefore, a common requirement is to consider dependence measures that are invariant to strictly increasing transformations of the marginal variables. The copula-based dependence measures fulfill the above postulate and therefore are scale invariant. Several real-valued measures of such type have been studied for a long time. For a nice overview see \cite{r42}, and \cite{r10}. 
However, nowadays there is strong evidence that attempting to express a complex dependence structure via a single number is hopeless. \cite{r50} admitted that such a conclusion was clearly spelled out as early as in the celebrated book by \cite{r24}. 

To help understand the underlying dependence structure with aid of a procedure invariant under strictly increasing transformations of the marginal distributions, \cite{r14} proposed a rank-based graphical tool called a chi-plot.  The procedure is rather complicated and not easy to interpret. 
A simpler solution was introduced by \cite{r15}. Their display is called a Kendall plot and refers to the idea of a Q-Q plot. It also is based on ranks and is more directly related to the underlying copula function than the chi-plot. See also \cite{r51} for further development of this idea.
\cite{r26,r27} has proposed a simple dependence measure $q$ which is explicitly based on the copula function. The measure aggregates some Fourier coefficients of the copula in some special non-orthogonal basis. 
Expansions in the Hilbert space using systems of functions which are not mutually orthogonal appear in the literature under the label ``quasiorthogonal expansions''; cf. \cite{r62}. The above mentioned Fourier coefficients can also be interpreted in terms of some local correlations; cf. \cite{r29}, p. 39, and \cite{r28}, p. 361. For further interpretation see Remark \ref{rem1}, below. Here we mention only that the measure $q$ can be seen to be naturally related to the cross-quantilogram, a notion with growing importance in econometry.

The remainder of the article is organized as follows: Section \ref{S2} recalls the definition of the measure $q$ and its properties, shows some links of $q$ to other existing notions, and illustrates the shape of the measure in a series of interesting bivariate models considered in recent literature on independence testing. Section \ref{S3} introduces a new useful estimator of $q$. In Section \ref{S4}, we present three new test statistics pertaining to the proposed estimator of $q$. Two test statistics are the supremum-norm and integral-norm of the estimated $q$, respectively. The third solution exploits the minimum $p$-value principle. 
We also report there basic theoretical results on the new tests. In Section \ref{S5}, we provide a comparative study on power analysis for the new solutions and some competitors which have been already introduced in the literature on the subject. Section \ref{S6} contains a study of a real data example. Proofs of all results are relegated to Supplementary Materials, which are located at final part of the manuscript. In the Supplementary Materials we include also some comments on our implementation of new statistics, and provide related C codes.     
\vspace{-0.6cm}
\section{A Copula-Based Measure of Dependence: the Quantile Dependence Function}
\label{S2}
\vspace{-0.3cm}
\subsection{Definition and Properties}
\label{S2.1}
Consider a pair of random variables $X$ and $Y$ with bivariate cumulative distribution function $H$ and with continuous margins $F$ and $G$, respectively. Then there exists a unique copula $C$ such that $C(u,v)=H(F^{-1}(u),G^{-1}(v)),\;(v,v) \in [0,1]^2$. Obviously, $F^{-1}(u)$ and $G^{-1}(v)$, appearing in this formula, are the $u$-quantile and $v$-quantile of the respective marginal distribution functions. Set
\vspace{-0.2cm}
\begin{equation}\label{eq1}
q(u,v)=q_C(u,v)=\frac{C(u,v)-uv}{\sqrt{uv(1-u)(1-v)}}, \;\;\;\;(u,v) \in (0,1)^2. 
\end{equation}
We shall call $q$ {\it the quantile dependence function}.
By (1), the measure $q$ attributes the copula $C$ to the continuous function on $(0,1)^2$. 
As stated and justified in \cite{r26,r27}, the measure $q$ fulfills natural postulates, motivated by the axioms formulated in \cite{r42} and updated in \cite{r12}. For convenience, we recall here the basic properties of $q$. Below, we set
\vspace{-0.4cm} 
\begin{equation}\label{eq2}
w(u,v)=1/\sqrt{uv(1-u)(1-v)}.
\vspace{-0.4cm} 
\end{equation}
\begin{pro}\label{pro1} 
{ The quantile dependence function $q$, given by (\ref{eq1}),  has the following properties.
\begin{enumerate}
\vspace{-0.2cm}
\item $-1 \leq q(u,v) \leq 1$ for all $(u,v) \in (0,1)^2$.
\vspace{-0.4cm}
\item  By the Fr\'echet-Hoeffding bounds for copulas, the property 1 can be further sharpened to 
$
B_o(u,v) \leq q(u,v) \leq B^o(u,v),\;\;(u,v) \in (0,1)^2,
$
where $B_o(u,v)=w(u,v) \times$  $[\max\{u+v-1,0\} -uv]$ and $B^o(u,v)=w(u,v)[\min\{u,v\} -uv]$. 
\vspace{-0.4cm}
\item {$q$ is maximal (minimal) if and only if $Y=f(X)$ and $f$ is strictly increasing (decreasing) a.s. on the range of $X$.} \vspace{-0.4cm}
\item {$q(u,v)\equiv 0$ if and only if $X$ and $Y$ are independent.}
\vspace{-0.4cm}
\item The equation $q(u,v)\equiv c$, $c$ a constant, can hold true if and only if $c=0$.
\vspace{-0.4cm}
\item $q$ is non-negative (non-positive) if and only if $(X,Y)$ are positively (negatively) quadrant-dependent.
\vspace{-0.4cm}
\item $q$ is invariant under transformations which are strictly increasing a.s. on ranges of $X$ and $Y$, respectively.
\vspace{-0.4cm}
\item If $X$ and $Y$ are transformed by strictly decreasing a.s. functions, then $q(u,v)$ is transformed to $q(1-u,1-v)$.
\vspace{-0.4cm}
\item If $f$ and $g$ are strictly decreasing a.s. on ranges of $X$ and $Y$, respectively, then $q$'s for the pairs $(f(X),Y)$ and $(X,g(Y))$ take the forms $-q(1-u,v)$ and $-q(u,1-v)$, accordingly.
\vspace{-1.2cm}
\item $q$ respects concordance ordering, i.e. for cdf's $H_1$ and $H_2$ with the same marginals and corresponding copulas $C_1$ and $C_2$, $H_1(x,y) \leq H_2(x,y)$ for all $(x,y) \in \mathbb{R}^2$ implies 
$q_{C_1}(u,v) \leq q_{C_2}(u,v)$ for all $(u,v) \in (0,1)^2$.
\vspace{-0.4cm}
\item If $(X,Y)$ and $(X_n,Y_n),\;n=1,2,\ldots,$ are pairs of random variables with joint cdf's $H$ and ${\check{H}}_n$, and the corresponding copulas $C$ and ${\check{C}}_n$, respectively, then weak convergence of $\{\check{H}_n\}$ to $H$ implies $q_{\check{C}_n}(u,v) \to q_C(u,v)$ for each $(u,v) \in (0,1)^2$.
\end{enumerate}
}
\end{pro}

\vspace{-0.4cm}
\begin{rem}\label{rem1}
{ 
The measure 
$q(u,v)$ can be explicitly related to some tail-dependence indices:
 lower tail-dependence coefficient $\lambda_L=\lim_{ u \searrow 0}\frac{C(u,u)}{u}=\lim_{ u \searrow 0}q(u,u)$ and
 the coefficient of upper tail dependence $\lambda_U = \lim_{ u \nearrow 1} \frac{1-2u + C(u,u)}{1-u}$ $= \lim_{ u \nearrow 1} q(u,u)$, which are defined in the situations when  the respective limits exist;
 coefficients of tail dependence: $\tau^{UU}(u)=P(U > u|V > u),\;$ $\tau^{LL}(u) = P(U < 1-u | V < 1-u),$ $\tau^{LU}(u) = P(U < 1-u | V > u) $\; and
$\tau^{UL}=P(U>u|V<1-u)$. In particular, on the diagonal, $q(u,u)=\tau^{UU}(u) + \tau^{LL}(1-u) -1$ while on anti-diagonal $q(u,1-u)=1-\tau^{UL}(u)-\tau^{LU}(1-u)$.
See \cite{r44}, and \cite{r6}, respectively. 

The measure $q$ is weak-equitable and weakly-robust-equitable in the sense of Definitions 2 and 4 in \cite{r11}, accordingly.

For radially symmetric copulas $(C(u,v)=\underline{C}(1-u,1-v);$ where $ \underline{C}(u,v)=P(U>u,V>v))$ it holds that $q(u,u)= \beta ((u,u),(1,1))$, where $\beta((u_1,u_2),(v_1,v_2))$ stands for generalized Blomqvist's measure of concordance considered in \cite{r41}. 

Finally note that $q$ coincides with the cross-quantilogram applied to $(X,Y)$. Therefore, using precise terminology, we can say that $q(u,v)$ is the cross-correlation of the respective quantile hits.
The cross-quantilogram for time series was, somewhat in passing,  introduced on p. 261 in \cite{r30} in the context of predictability studies. In \cite{r17} the idea has received considerable attention and development. }
\end{rem} 
\vspace{-0.5cm}
\subsection{Graphical illustration}
\label{S2.2}

\begin{figure}[!ht]
\centering
\includegraphics[scale=0.74]{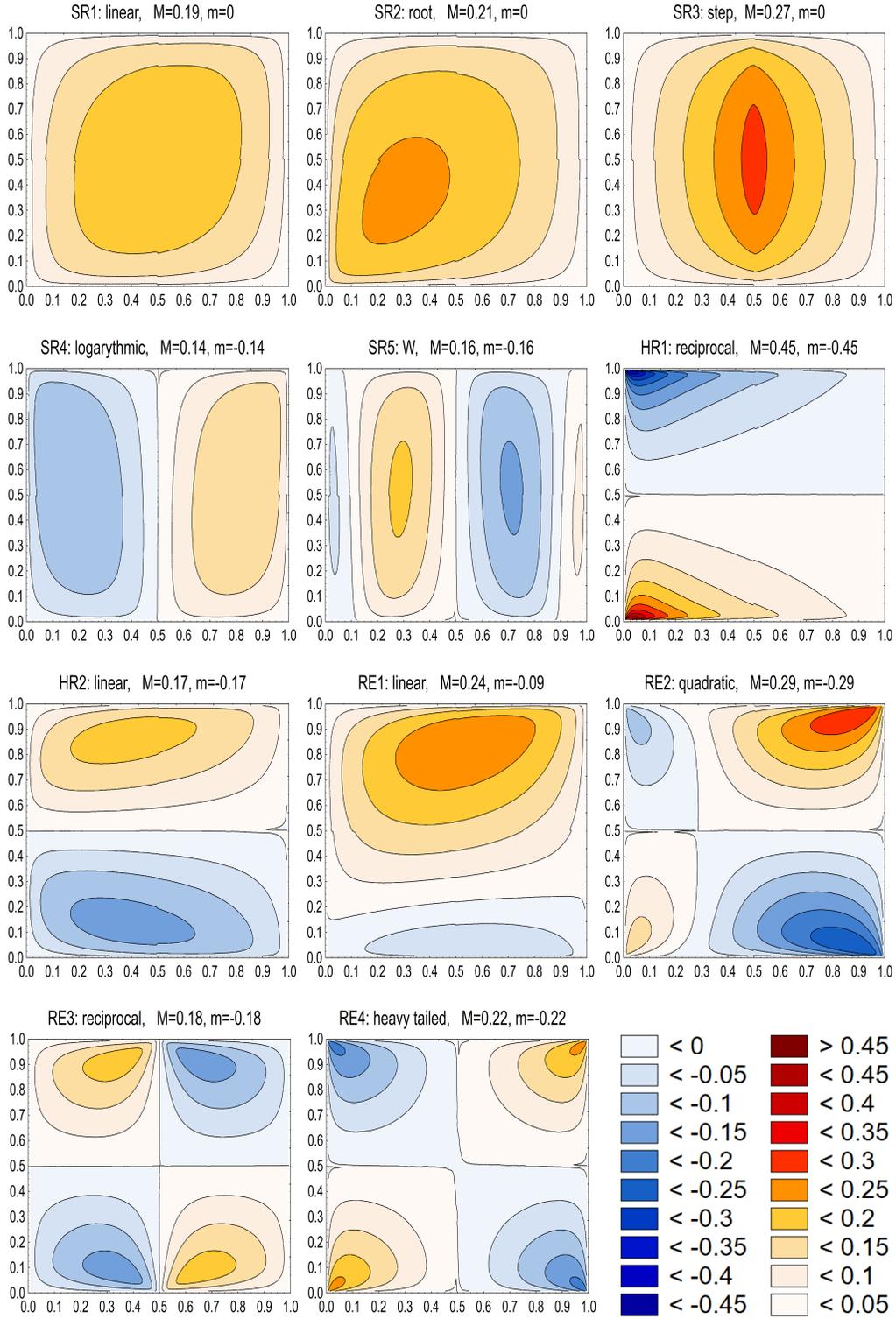} 
\vspace{-0.1cm}
\caption{Shapes of $q$. Regression models.}\label{fig1}
\end{figure}

\begin{figure}[!ht]
\centering
\includegraphics[scale=0.74]{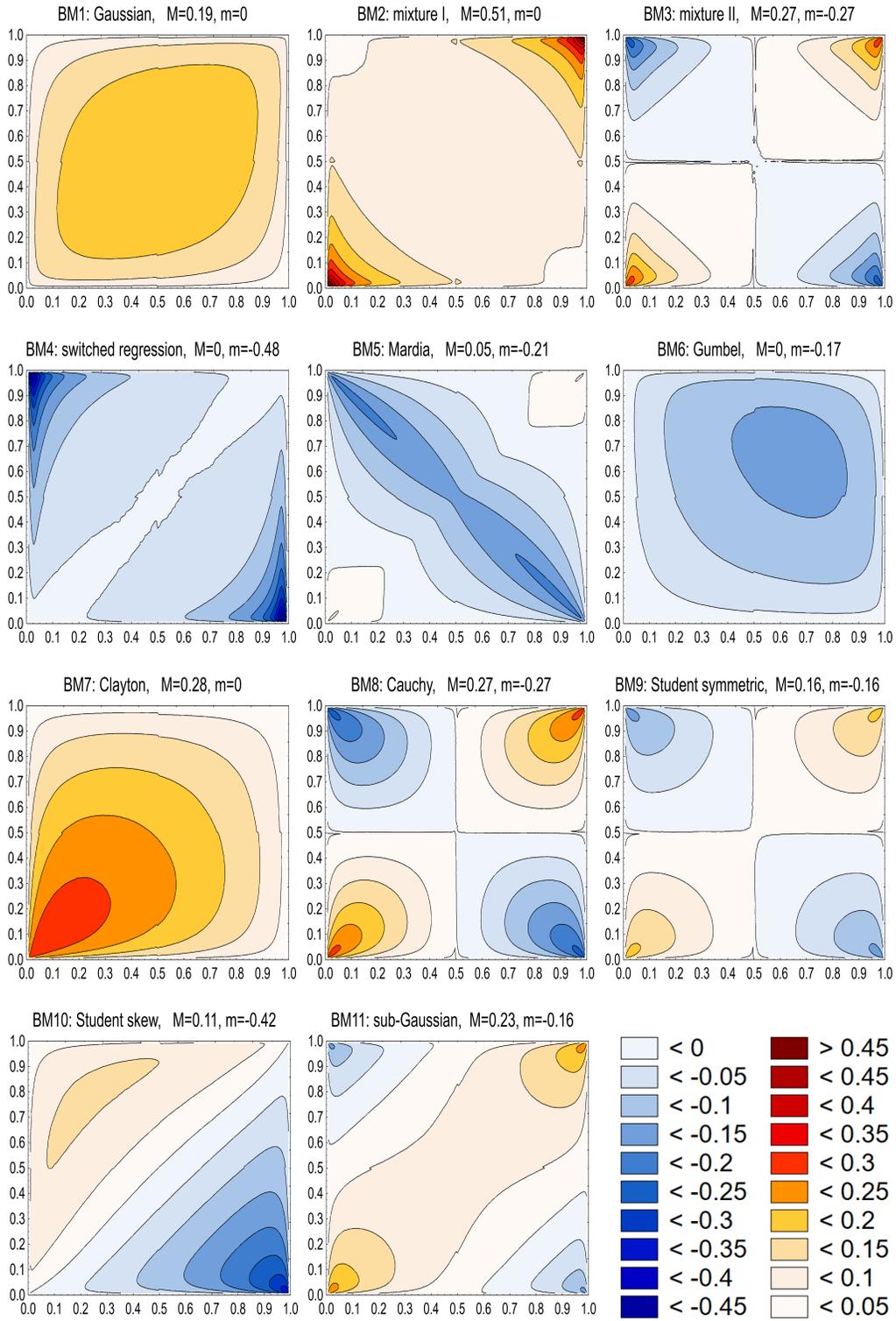} 
\vspace{-0.1cm}
\caption{Shapes of $q$. Classical bivariate models.}\label{fig2}
\end{figure}

Given a  formula for $C$, the quantile dependence function $q$ can be graphically presented in many convenient ways. However, in the next section we introduce a new estimate $Q_n^*$ of $q$ and here we simply display the corresponding (averaged) values of the estimate for large $n$, and for several interesting forms of dependence of $X$ and $Y$.  To be specific, in Figures \ref{fig1} and \ref{fig2} we took $n=1000$ and calculated the averages over 10 000 MC runs. 
This gives quite precise information on the shape of $q$.

To see how the measure $q$ reflects different forms of dependence and to study empirical powers of new tests proposed in Section \ref{S4}, we have considered a wide spectrum of models. Some of them are classical ones, a considerable portion has been recently introduced in different simulation experiments in some related papers, a few of the models have been defined just for this study. Below we only show the sources of some recently introduced models. Classical ones were used in numerous earlier simulation studies. Throughout the paper ${\bf 1}(A)$ stands for the indicator of the set $A$. The list of models is as follows.\\
\noindent
{\bf Simple Regression}:
\begin{itemize}
\vspace{-0.4cm}
\item[SR1:] {\it linear}\ \ \  %(P5) 
$\;\;Y=2+X+\epsilon,\ \;X\sim U[0,1],\ \;\epsilon \sim N(0,1)$;\ \ \cite{r49};
\vspace{-0.4cm}
\item[SR2:] {\it root} \ \ %(reg $X^{1/4}$) 
$\;\;Y=X^{1/4} +\epsilon,\ \;X\sim U[0,1],\ \;\epsilon \sim N(0,0.25)$;\ \ \cite{r45},\ \  \cite{r37};
\vspace{-0.4cm}
\item[SR3:] {\it step} \ %(reg step) 
$\;\;Y={\bf 1}(X\leq0.5) + \epsilon,\;X\sim U[0,1],\ \epsilon \sim N(0,2)$;  \cite{r45},  \cite{r37};
\vspace{-0.4cm}
\item[SR4:] {\it logarithmic} %(P3) 
$\;\;Y=\log(1+|X|) + \epsilon,\;X \sim N(0,1);\epsilon \sim N(0,1)$;  \cite{r49};
\vspace{-0.4cm}
\item[SR5:] {\it W} %(reg W) 
$\;\;Y=4[(2X-1)^2 -0.5]^2+\epsilon,\;X\sim U[0,1],\;\epsilon \sim N(0,0.5)$;  \cite{r10}.
\end{itemize}
\hspace{-0.08cm}\textbf{Heterosceadestic Regression}:
\begin{itemize}
\vspace{-0.4cm}
\item[HR1:] {\it  reciprocal} 
$\;\;Y=\sigma(X)\epsilon,\;$ $X$ has exponential distribution with $\lambda=0.1$, $\epsilon \sim N(0,1),\;\sigma(X)=\sqrt{1+ 1/ X^2}$;
\vspace{-0.4cm}
\item[HR2:] {\it linear  } 
$\;\;Y=\sigma(X)\epsilon,\;$ $X \sim U[1,16],\;$ $\epsilon \sim N(0,1),\;\sigma(X)=\sqrt X$.
\end{itemize}
\hspace{-0.05cm}{\bf Random-Effect-Type Models}:
\begin{itemize}
\vspace{-0.4cm}
\item[RE1:] {\it linear}\  %(P6) 
$\;\;Y=2+X+\epsilon_M X + \epsilon_A,\ \;X\sim U[0,1],\ \;\epsilon_M \sim N(0,4),\ \;\epsilon_A \sim N(0,1)$,\ \  \cite{r49};
\vspace{-0.4cm}
\item[RE2:] {\it quadratic} %(P1) 
$\;\;Y=\epsilon_M(2+X+X^2)+\epsilon_A,\;\epsilon_M \sim N(0,1),\;\epsilon_A \sim N(0,1)$, \cite{r50};
\vspace{-0.4cm}
\item[RE3:] {\it reciprocal} %(P2) 
$\;\;Y=\epsilon_M X^{-1}+\epsilon_A,\;\epsilon_M \sim N(0,1),\;\epsilon_A \sim N(0,1)$, \cite{r50};
\vspace{-0.4cm}
\item[RE4:] {\it heavy tailed} %(P8) 
$\;(X_0,Y_0)$ is bivariate Cauchy. Define $X=X_0, Y=\epsilon_M Y_0 +\epsilon_A$, $\epsilon_M \sim N(0,1),\;\epsilon_A \sim N(0,1)$, Supplemental Material for \cite{r50}.
\end{itemize}
\hspace{-0.12cm}{\bf  Bivariate Models}: 
\begin{itemize}
\vspace{-0.4cm}
\item[BM1:] {\it Gaussian}\;\; 
bivariate normal distribution with  $\rho=0.3$;
\vspace{-0.4cm}
\item[BM2:] {\it mixture I}\;\; 
the mixture (0.1)(standard bivariate Gaussian) + (0.9)(bivariate Gaussian with mean 0, variances 6 and the covariance 5);
%(MixNorm($p=0.1, \sigma^2=6, Cov(X,Y)=5$));
\vspace{-0.4cm}
\item[BM3:] {\it mixture II}\;\; 
the mixture (0.3)(bivariate Cauchy) + (0.7)(standard bivariate Gaussian);
\vspace{-0.4cm}
\item[BM4:] {\it switched regression}\;\; 
$Y=\mu(X) + \epsilon,\;\epsilon \sim N(0,1),\;\mu(X)=0$ for $|X|\leq 1.96$ and $\mu(X)=-X$ otherwise;
\vspace{-0.4cm}
\item[BM5:] {\it Mardia}$\;\;$ 
Mardia family of copulas with $\theta=-0.55$;% \cite{r31}; 
\vspace{-0.4cm}
\item[BM6:] {\it Gumbel}$\;\;$ 
Gumbel bivariate distribution with $\theta=0.5$;
\vspace{-0.4cm}
\item[BM7:] {\it Clayton}\;\; 
Clayton model with $\theta=0.5$;
\vspace{-0.4cm}
\item[BM8:] {\it Cauchy}\;\; 
bivariate Cauchy distribution; 
\vspace{-0.4cm}
\item[BM9:] {\it Student symmetric}\;\; 
symmetric Student's distribution with 2 degrees of freedom;
\vspace{-0.4cm}
\item[BM10:] {\it Student skew}\;\; 
skew bivariate Student's distribution with 5 degrees of freedom and parameters $(0.3,0.7,-0.7)$;%, \cite{r9};
\vspace{-0.4cm}
\item[BM11:] {\it sub-Gaussian}\;\;
bivariate sub-Gaussian distribution with parameters $(0.1,1.5)$, \cite{r23}. % \cite{r39},
\end{itemize}

%\newpage

Each panel in Figures 1 and 2 only has a short label related to the above detailed description. Moreover, to increase readability, each label also contains two numbers $M$ and $m$ which are the maximal and minimal values, respectively, of the corresponding $q(u,v)$ for $(u,v) \in (0,1)^2$.
\vspace{-0.3cm} 
\section{Symmetrized estimate of $q$}
\label{S3}
\vspace{-0.2cm} 
\subsection{Motivation}
\label{S3.1}

Let $(X_1,Y_1),...,(X_n,Y_n)$ be independent and identically distributed random vectors drawn from a bivariate distribution function $H$ with continuous marginals $F$ and $G$. Furthermore, let $R_i$ denote the rank of $X_i,\;i=1,...,n,$ in the sample $X_1,...,X_n$, while $S_i$ stands for the rank of $Y_i,\;i=1,...,n,$ in the sample $Y_1,...,Y_n$. 

A natural way of estimating $q$, given by (1), is the plug-in method. There are several estimators of $C$ available. The first proposal presumably goes back to \cite{r38}, see pp. 6 and 12,  and has the form
$$
 E_n(u,v)=\frac{1}{n}\sum_{i=1}^n {\bf 1}\Bigl(\frac{R_i}{n} \leq u,\frac{S_i}{n} \leq v\Bigr).
$$ 
Monte Carlo simulations show that it is better to use
\begin{equation}\label{eq3}
 C_n(u,v)=\frac{1}{n}\sum_{i=1}^n {\bf 1}\Bigl(\frac{R_i}{n+1} \leq u,\frac{S_i}{n+1} \leq v\Bigr).
\end{equation}
Under the above assumptions, $C_n$ and $E_n$ differ by $O(1/n)$ almost surely. % cf. (B.1) in Genest and Segers (2009), for example.  
 Alternatively, one can include the continuity correction in $E_n(u,v)$. Another option are kernel based smooth versions of $E_n(u,v)$ or $ C_n(u,v)$; cf. \cite{r13},  \cite{r33} and the references therein. A different smoothed variant has been proposed by \cite{r40}. They introduced the Bernstein copula estimator, which has been further studied by \cite{r21}, and \cite{r43}, among others.

It should be emphasized, however, that our ultimate goal is not the estimation of the copula itself, but a construction of some tests of independence based on the corresponding estimator of $q$. In such an application, excessive smoothing of the underlying parameter, i.e. the copula function, is often not profitable to the power of the resulting test.
On the other hand, we shall  consider test statistics as functionals of the process $w(u,v)\{\check{C}_n(u,v)-uv\}$,  where $\check{C}_n$ is an estimate of the copula, while the weight is given by (\ref{eq2}).
Therefore, we have to be very careful about the behavior of the quantity $\{\check{C}_n(u,v)-uv\}$ near the edges of $[0,1]^2$. In this sense, $C_n$ is not convenient for our purposes. It can be seen that the expression $Q_n(u,v)=w(u,v)\{C_n(u,v)- uv\}$ takes on very large (absolute) values near the points (0,1), (1,0) and (1,1).
In contrast, empirical behavior of $Q_n(u,v)$ near (0,0) is satisfactory. Therefore, 
to estimate the numerator of $q(u,v)$, i.e. the function 
$$
N(u,v)=C(u,v)-uv,
$$ 
say, we propose and we shall apply a symmetrized variant of the random function 
$$
N_n(u,v)=C_n(u,v)-uv,
$$ 
which exploits the useful behavior of $Q_n(u,v)$ in the first quadrant.
\vspace{-0.1cm} 
\subsection{Symmetrization}\label{S3.2}
\vspace{-0.1cm}
To define the symmetrization, note that the numerator $N(u,v)$ of $q(u,v)$, for every $(u,v) \in [0,1]^2$, can be rewritten in the following four forms
\begin{eqnarray}
N(u,v)&=&N^{(1)}(u,v)= P(F(X)\leq u,G(Y)\leq v) -uv \nonumber \\
&=&N^{(2)}(u,v)= u(1-v)-P(F(X)\leq u, G(Y) > v) \nonumber \\
&=&N^{(3)}(u,v)= P(F(X)>u,G(Y)>v)-(1-u)(1-v), \nonumber \\
&=&N^{(4)}(u,v)= (1-u)v -P(F(X)>u,G(Y)\leq v). \nonumber 
\end{eqnarray}
\noindent
Therefore, similarly as in (\ref{eq3}), we can consider the following variants of estimators of $N(u,v)$.
\begin{eqnarray}
N^{(1)}_n(u,v)&=& \frac{1}{n}\sum_{i=1}^n {\bf 1}\bigl(\frac{R_i}{n+1} \leq u, \frac{S_i}{n+1} \leq v\bigr) - uv, \nonumber \\
N^{(2)}_n(u,v)&=&  u(1-v) -\frac{1}{n}\sum_{i=1}^n {\bf 1}\bigl(\frac{R_i}{n+1} \leq u,\frac{S_i}{n+1}>v\bigr), \nonumber \\
N^{(3)}_n(u,v)&=& \frac{1}{n}\sum_{i=1}^n {\bf 1}\bigl(\frac{R_i}{n+1} > u, \frac{S_i}{n+1} > v\bigr) - (1-u)(1-v), \nonumber \\
N^{(4)}_n(u,v)&=& (1-u)v -\frac{1}{n}\sum_{i=1}^n {\bf 1}\bigl(\frac{R_i}{n+1} > u,\frac{S_i}{n+1}\leq v\bigr). \nonumber 
\end{eqnarray}
\noindent
This leads us to the symmetrized estimator of $N(u,v),\;(u,v) \in [0,1]^2,$ given by
\begin{eqnarray}
N_n^*(u,v)&=&N^{(1)}_n(u,v){\bf 1}(u \leq \frac{1}{2},v \leq \frac{1}{2}) + N^{(2)}_n(u,v){\bf 1}(u \leq \frac{1}{2},v > \frac{1}{2}) \nonumber \\
& &+N^{(3)}_n(u,v){\bf 1}(u > \frac{1}{2},v > \frac{1}{2}) + N^{(4)}_n(u,v){\bf 1}(u > \frac{1}{2},v \leq \frac{1}{2}).\label{eq4} 
\end{eqnarray}
\noindent
Note that for any $(u,v) \in [0,1]^2$ it holds that
\begin{equation}\label{eq5}
N_n^*(0,v)=N_n^*(u,0)= N_n^*(u,1)=N_n^*(1,v)=0.
\end{equation}

We shall call $\{\sqrt n N_n^*(u,v),\;(u,v) \in [0,1]^2\}$ the symmetrized version of the  process $\{\sqrt n [C_n(u,v)-uv],\;(u,v) \in [0,1]^2\}$.  Finally, the initial and symmetrized estimators of $q(u,v)$ are given by
\begin{equation}\label{eq6}
Q_n(u,v)=w(u,v)N_n(u,v)\;\;\;\mbox{and}\;\;\;Q_n^*(u,v)=w(u,v)N_n^*(u,v),
%\frac{N_n(u,v)}{\sqrt{uv(1-u)(1-v) }}\;\;\;\mbox{and}\;\;\;Q_n^*(u,v)=\frac{N_n^*(u,v)}{\sqrt{uv(1-u)(1-v) }},
\end{equation}
respectively. 
Note that outside $[1/{(n+1)},1-1/{(n+1)}]^2$ the estimator $Q_n^*(u,v)$ is deterministic, bounded, and it tends to 0 when its arguments approach the edges of $[0,1]^2$. This makes a great difference in finite sample behavior in comparison with $Q_n(u,v)$, given in (\ref{eq6}) as well. 

We shall also consider a smoothed variant of $Q_n^*$. To be specific, for some $s \geq 0$ we shall use
\begin{equation}\label{eq7}
Q_{n,s}^*(u,v) = \Bigl(\frac {2s}{n+1}\Bigr)^{-2} \int_{u-\frac{s}{n+1}}^{u+\frac{s}{n+1}} \int_{v-\frac{s}{n+1}}^{v+\frac{s}{n+1}} Q_n^*(x,y)dy dx.
\end{equation}

The three above defined estimators of $q$ have the following useful property:

\begin{pro}\label{pro2}
{ Let $(X_1,Y_1),...,(X_n,Y_n)$ be independent and identically distributed random vectors 
drawn from a bivariate population $(X,Y)$ obeying a joint distribution function $H$ with continuous marginals $F$ and $G$. Under the independence of $X$ and $Y$, given $(u,v) \in (0,1)^2$, it holds that
\begin{equation}\label{eq8}
\sqrt n Q_n(u,v) \stackrel{D}{\rightarrow} N(0,1),
\;\;\;\sqrt n Q_n^*(u,v) \stackrel{D}{\rightarrow} N(0,1)\;\;\;\mbox{and}\;\;\;\sqrt n Q_{n,s}^*(u,v) \stackrel{D}{\rightarrow} N(0,1),
\end{equation}
where $\stackrel{D}{\rightarrow}$ stands for convergence in distribution.}
\end{pro}

This, in particular,  makes it possible to immediately see on the graphs of the estimates of $q(u,v)$ some evidence indicating in which part of the population and to which extent (at least roughly) the independence is invalidated. Formal tests, based on some functionals of these estimators, are defined in Section \ref{S4}.

\section{New weighted test statistics and their properties}
\label{S4}

For continuous random variables, testing for independence is equivalent to verification if the true copula is equal to the independence copula. Therefore, we consider the null hypothesis of the form 
$$
{\bf H_0} : C(u,v)=uv,\;\;\;\mbox{for}\;\;\;(u,v) \in [0,1]^2
$$
and study test statistics defined as some functionals of the estimated difference $C(u,v)-uv$.

\subsection{Integral statistics}
\label{S4.1}

Weighted integral copula-based statistics of the form 
$$
n\int_0^1\int_0^1 u^{2\gamma}v^{2\delta}\{E_n(u,v)-uv\}^2 dudv,
$$
where $\gamma>-1/2$ and $\delta >-1/2$, were studied in \cite{r7}. Recently, \cite{r4} have introduced the independence statistic which in the two dimensional case reads as 
$$
n \int_0^1\int_0^1 \{g(u,v)\}^{-\gamma}\{C_n^{\beta}(u,v)-uv\}^2dudv,
$$
where $C_n^{\beta}$ is the empirical beta copula - a particular case of the empirical Bernstein copula, $\gamma \in [0,2)$, and $g(u,v)=\min\{u,v,1-\min(u,v)\}$. 

We shall consider the (standardized) $L_r$-norm of $Q_n^*(u,v)$ 
\begin{equation}\label{eq9}
{\cal L}_{r,n}^* = \sqrt n \Bigl\{\int_0^1 \int_0^1 \big|Q_n^*(u,v)\big|^r du dv\Bigr\}^{1/r}.
\end{equation}
In view of the form of $Q_n^*$, cf. (6), ${\cal L}_{r,n}^*$ is a weighted integral-type statistic for the symmetrized version of the  process 
$\{\sqrt n [C_n(u,v)-uv]\}$, where the weight is given by (2). For relatively small $r$ one could find a closed expression for ${\cal L}_{r,n}^*$. In general, for reasonably large values of $n$, it suffices to approximate the value of 
${\cal L}_{r,n}^*$ by $\frac{\sqrt n}{(n+1)^{2}}\bigl\{\sum_{i=1}^n \sum_{j=1}^n \big|Q_n^*(\frac{i+0.5}{n+1},\frac{j+0.5}{n+1})\big|^r\bigr\}^{1/r}$.

To achieve consistency of statistics like (\ref{eq9}), using some existing results on the classical empirical copula process, we have to slightly modify the integration area in (\ref{eq9}) around the vertices of $[0,1]^2$. For an illustration, given $\epsilon > 0$, we shall consider
\begin{equation}\label{eq10}
A({\epsilon})= [0,1]\setminus \left\{[0,\epsilon]^2 \cup [1-\epsilon,1]\times[0,\epsilon] \cup [1-\epsilon,1]^2 \cup [0,\epsilon]\times[1-\epsilon,1]\right\}
\end{equation}
and  the modified variant of ${\cal L}_{r,n}^*$ given by
$$
{\cal L}_{\epsilon,r,n}^*= \sqrt n \Bigl\{\int_{A({\epsilon})} |Q_n^*(u,v)|^r dudv\Bigr\}^{1/r}.
$$
   
To derive the consistency of ${\cal L}_{\epsilon,r,n}^*$, some smoothness assumptions have to be imposed on the underlying copula $C$. The following non-restrictive requirements have been formulated in \cite{r56} and turn out to be sufficient in many practically important situations. For more details, see \cite{r56}, and \cite{r4}.%\cite{r3}  

\begin{as}\label{as0} \ \vspace{0.01cm}
\begin{enumerate}
%{\it
\item $\frac{\partial}{\partial u}C(u,v)$ and $\frac{\partial}{\partial v}C(u,v)$ exist and are continuous on $R_1=(0,1)\times[0,1]$ and $R_2=[0,1]\times(0,1)$, respectively;
\item $\frac{{\partial}^2}{\partial u \partial v} C(u,v)$ exists and is continuous on $R_1 \cap R_2$;
\item There exists a constant $K >0$ such that 
$\;\;\Bigl|\frac{{\partial}^2}{\partial u \partial v} C(u,v)\Bigr| \leq K \min \Bigl\{ \frac{1}{u(1-u)},\frac{1}{v(1-v)}\Bigr\},\;$ for $(u,v) \in R_1 \cap R_2$.
%}
\end{enumerate}
\end{as}

\begin{pro}\label{pro3}
{Assume that the underlying copula $C$ satisfies the Assumption \ref{as0}. Consider any $r>2$ and a positive $\epsilon=\epsilon_n$. Suppose that $\epsilon_n  \to 0$ in such a way that $\epsilon_n n^{1/(r+2)} \to \infty$, as $n \to \infty$.
Then, under the alternative corresponding to $C$,  the test rejecting ${\bf H_0}$ for large values of ${\cal L}_{\epsilon,r,n}^*$ is consistent.}
\end{pro}

\subsection{Minimum $p$-value statistics}
\label{S4.2}

A test rejecting ${\bf H_0}$ for large values of the distance ${\cal L}_{\epsilon,r,n}^*$ is expected to be especially sensitive to alternatives shifting the probability mass  towards the edges of $[0,1]^2$. In contrast, the statistic proposed in \cite{r18} proves to be very efficient in detecting some noisy functional relationships. Therefore, it seems useful to propose a procedure which combines advantages of both solutions. Our approach to this question is via combining $p$-values. For this purpose denote by ${\cal T}_n$ the rank variant of the statistic, based on pairwise distances,  introduced in \cite{r18}. 
%In our simulation study we use the notation HHG for this solution.
Given the sample $(X_1,Y_1),...,(X_n,Y_n)$, let $p^{(1)}_n$ denote $p$-value of ${\cal T}_n$ and let $p^{(2)}_n$ be respective $p$-value of ${\cal L}_{\epsilon,r,n}^*$. 

We propose to reject ${\bf H_0}$ for small values of 
\begin{equation}\label{eq11}
{\cal M}_n^* = \min \bigl\{p^{(1)}_n,p^{(2)}_n\bigr\}.
\end{equation}
The idea of a minimum $p$-value statistic goes back to Tippett \cite{r47}.

\subsection{Supremum-type solutions}
\label{S4.3}

Another classical approach to measuring departures from ${\bf H_0}$ is by taking a supremum norm of $Q_n^*$. More precisely, given $\kappa=\kappa_n \in (0,1/2)$ one can consider 
\begin{equation}\label{eq12}
{\cal D}_{\kappa,0,n}^*\sup_{(u,v) \in [\kappa,1-\kappa]^2} \sqrt n |Q_n^*(u,v)|. 
\end{equation}
Analogously to ${\cal L}_{r,n}^*$, ${\cal D}_{\kappa,0,n}^*$ can be interpreted as weighted sup-type statistic.
However, extensive simulations, which shall be partially reported in Section \ref{S5}, have shown that it is not necessarily a very powerful solution and some smoothing of $Q_n^*$ improves the finite sample behavior of such constructions. Therefore, we shall take  $Q_{n,s}^*(u,v)$ into account 
and introduce the corresponding class of supremum type statistics
\begin{equation}\label{eq13}
{\cal D}_{\kappa,s,n}^*=\sup_{(u,v) \in [\kappa,1-\kappa]^2} \sqrt n |Q_{n,s}^*(u,v)|. 
\end{equation}
Obviously, ${\cal D}_{\kappa,s,n}^*$ with $s=0$ coincides with the solution (\ref{eq12}). Supremum type statistics are particularly convenient for interpretation of obtained values of the selected estimator of the measure $q$ and at least rough, but practically immediate,  evaluation of the extent to which ${\bf H_0}$ is possibly invalidated.  

\begin{pro}\label{pro4}
Suppose that the requirement 1. of  Assumption \ref{as0} holds for $C$. Consider   positive $\kappa=\kappa_n,$ such that  $\;\kappa_n \to 0,$ and $\sqrt n \kappa_n \to \infty$,  as $n \to \infty.$ 
Then,  tests rejecting ${\bf H_0}$ for large values of ${\cal D}_{\kappa,0,n}^*$ and ${\cal D}_{\kappa,s,n}^*$, respectively, are consistent under the alternative $C$.
\end{pro}

\section{Simulated powers}
\label{S5}

We shall study the empirical behavior of the following statistics:
\begin{itemize}
\item the statistic {\it rank-dCov} of \cite{r46},  Section 4.3. We shall concisely denote this variant by $dCov$, as done in \cite{r19}, p. 17, as well;
\vspace{-0.4cm}
\item the rank based variant of the statistic introduced in \cite{r18}, denoted by HHG, similarly as proposed in  \cite{r19}, p. 17;
\vspace{-0.4cm}
\item the empirical likelihood ratio test $VT_n$, defined on  p. 160 of \cite{r49};
\vspace{-0.4cm}
\item ${\cal L}_{\epsilon,r,n}^*$ for two choices of $r: r=2,\;6$ and $\epsilon=0.01$; 
\vspace{-0.4cm}
\item ${\cal D}_{\kappa,s,n}^*$ for two  values of $s: s=0,\;4$ and $\kappa=0.025$; as usual in this area, the supremum in the Kolmogorov-Smirnov statistic was replaced by a maximum over a grid of $(n+1) \times (n+1)$ points.
\vspace{-0.4cm}
\item ${\cal M}_n^*$ using HHG and ${\cal L}_{\epsilon,6,n}^*$ with $\epsilon=0.01$. 
\end{itemize}

The outcomes of our Monte Carlo experiments, done for $n=100$ and the significance level $\alpha=0.05$, are collected in Table 1. Table 1 presents empirical powers under the 22 models introduced in Section \ref{S2}. %All code is in C.
%\smallskip
%\begin{table}[!h]
%\centering
%\caption{Critical values of selected statistics under $n=100$ and $\alpha=0.05; \kappa=0.025,\;\epsilon=0.01$. Based on 100,000 MC runs.}\label{tab1} 
%W TABELI WSTAWIC $\cal{D}_{\kappa,0,n}^*$ i $\cal{D}_{\kappa,4,n}^*$. uzaleznic L od $\epsilon$. TO SAMO W DRUGIEJ TABELI.
%\includegraphics[trim = 10mm 270mm 60mm 5mm, clip, scale=0.85]{Tab_1.pdf} 
%\end{table}
%\vspace{-0.5cm}
\begin{table}[!ht]
\centering
\caption{Empirical powers of selected statistics under $n=100$ and $\alpha=0.05; \kappa=0.025,\;\epsilon=0.01$. Based on 10,000 MC runs.}\label{tab2}
\includegraphics[trim = 10mm 140mm 40mm 5mm, clip, scale=0.85]{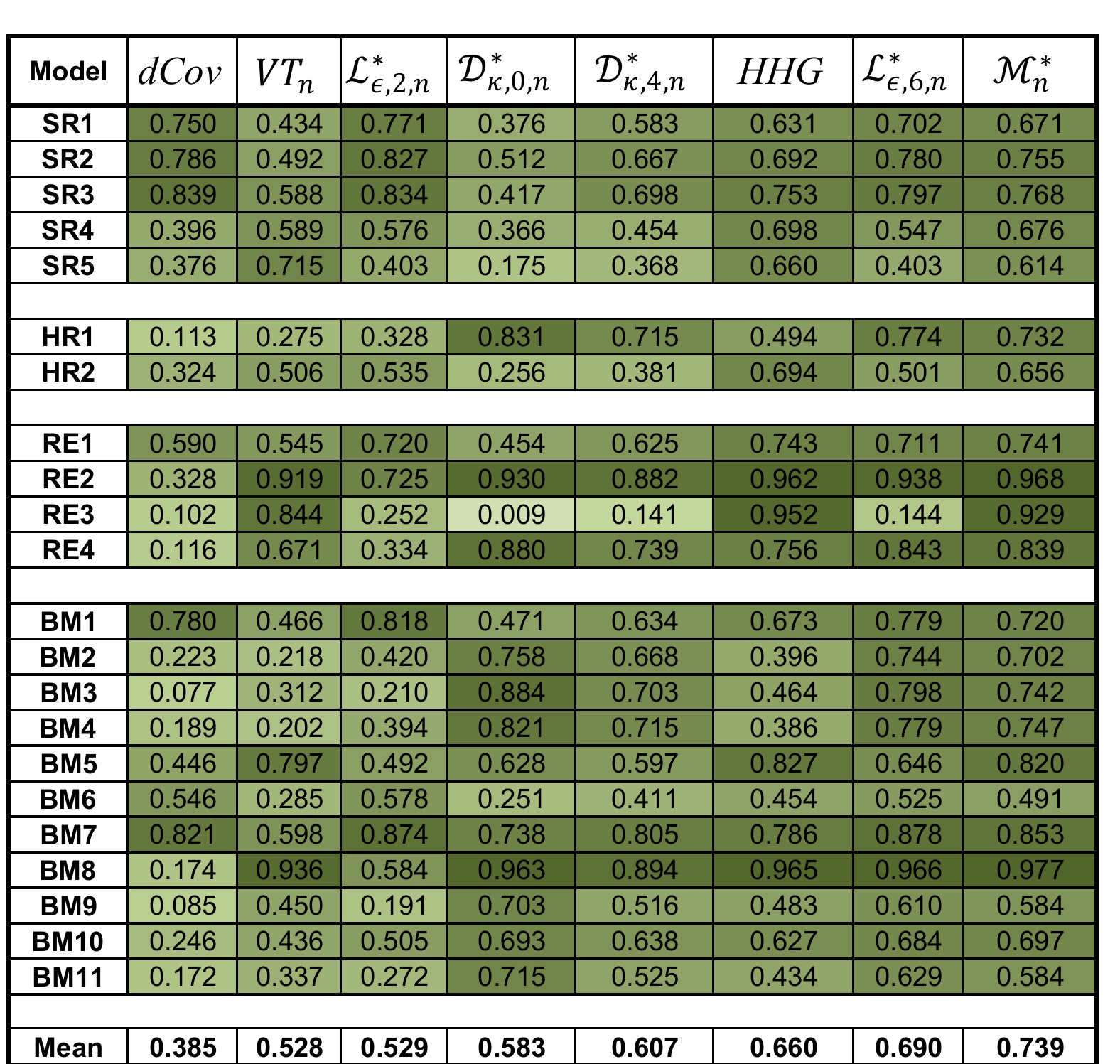} 
\end{table}

The simulation results  show that ${\cal D}_{\kappa,4,n}^*$ with $\kappa=0.025$ outperforms (in the average) the rank distance covariance and 
also slightly the empirical likelihood ratio test. The rank based statistic HHG is slightly more powerful (in the average) than ${\cal D}_{\kappa,4,n}^*$ while our second solution ${\cal L}_{\epsilon,6,n}^*$ with $\epsilon=0.01$ provides some improvement over the two last mentioned statistics. The most powerful solution turns out to be our third, ${\cal M}_n^*$, which combines the advantages of HHG (high power for regression models) with the benefits of ${\cal L}_{\epsilon,6,n}^*$ (sensitivity to heavy tails). It is also  worth noting that ${\cal D}_{\kappa,4,n}^*$ turns out to be much less sensitive to the choice of $\kappa$ than ${\cal D}_{\kappa,0,n}^*$. In particular, under our simulation scheme, for $\kappa < 0.025$ ${\cal D}_{\kappa,0,n}^*$ works much worse than its smoothed version ${\cal D}_{\kappa,4,n}^*$.

In view of the simulation results it appears that the class of statistics ${\cal L}_{\epsilon,r,n}^*$ shows considerable potential for further use. On the one hand, the integration process naturally smooths $Q_n^*$, and extra smoothing seems to be unnecessary. We have also investigated some smoother integrands in  $L_r$-type norm statistics, but they have not resulted in  more stable powers, and no average gain of power has been noticed. So, this is one advantage of such a solution over the supremum-type ones. On the other hand, it is well known that the $L_r$-norm, under $r \to \infty$, approximates the supremum norm. Therefore, the class is flexible and  enough reach. 
Needless to say, such smooth functionals of weighted empirical processes are also easier to analyze than weighted supremum-type statistics. Obviously, several questions arise in the context. The first one is the choice of $r$. In the course of the present simulation study we have simply inspected $r=2,...,8$ and decided for $r=6$. However, it is possible to study the problem more deeply and carefully, by calculating for example asymptotic relative efficiency of 
${\cal L}_{\epsilon,r,n}^*$ with respect to some standard, and investigating the classes of alternatives for which particular recommendations about $r$ are optimal. A recent paper by \cite{r20} provides tools that make it possible to answer to such a question. Obviously, solving such a problem is a fairly non-trivial task. 

Finally, note that the test statistic ${\cal M}_n^*$ performs very well. It has relatively simple structure which nicely reflects the advantages of its ingredients. The solution is similar in spirit to a much more complex one proposed in \cite{r19}. The latter combines $p$-values of chi-square-type tests over increasingly fine data-dependent sample space partitions.

\section{Application}
\label{S6}

We demonstrate our testing procedures on a data set of $n=230$ aircraft span (X) and speed (Y) data, on log scales, from years 1956-1984, collected by Saviotti, and reported and studied in \cite{r5}. Standard empirical correlation measures, such as Pearson's, Spearman's, and Blomqvist's rank statistics, applied to this data, do not invalidate independence, as their $p$-values are well above 0.7. This suggests that any dependence structure, if present at all, is likely to be nonlinear. \cite{r46}, and \cite{r18} have applied their tests to this data and have received very small $p$-values (less than 0.00001). In conclusion, the above evidence implies strong nonlinear dependence structure.

This data was also analyzed by \cite{r22}, and \cite{r2}. The two papers contain a presentation of some empirical dependence measures, defined  on $\mathbb{R}^2$. \cite{r22} have started with a local dependence function of \cite{r57} and proposed a so-called dependence map. The result of this approach suggests a rather complicated picture of the joint behavior of $X$ and $Y$; cf. their Figure \ref{fig2}. In turn, \cite{r2} have applied another local correlation concept, namely the local Gaussian correlation introduced in  \cite{r48}. The approaches by \cite{r22}, and \cite{r2},  are related to estimation and modeling bivariate densities on $\mathbb{R}^2$, respectively. The second solution is especially technically involved. The overall picture resulting from both implementations is very similar; cf. Figure 4 in \cite{r2}: on the plane there are three separated regions in which local dependence is positive, one region in which it is negative. There are also areas on the plane with no local dependence between log(span) and log(speed). 

\begin{figure}[!ht]
\centering
\includegraphics[trim = 30mm 90mm 30mm 80mm,scale=0.9]{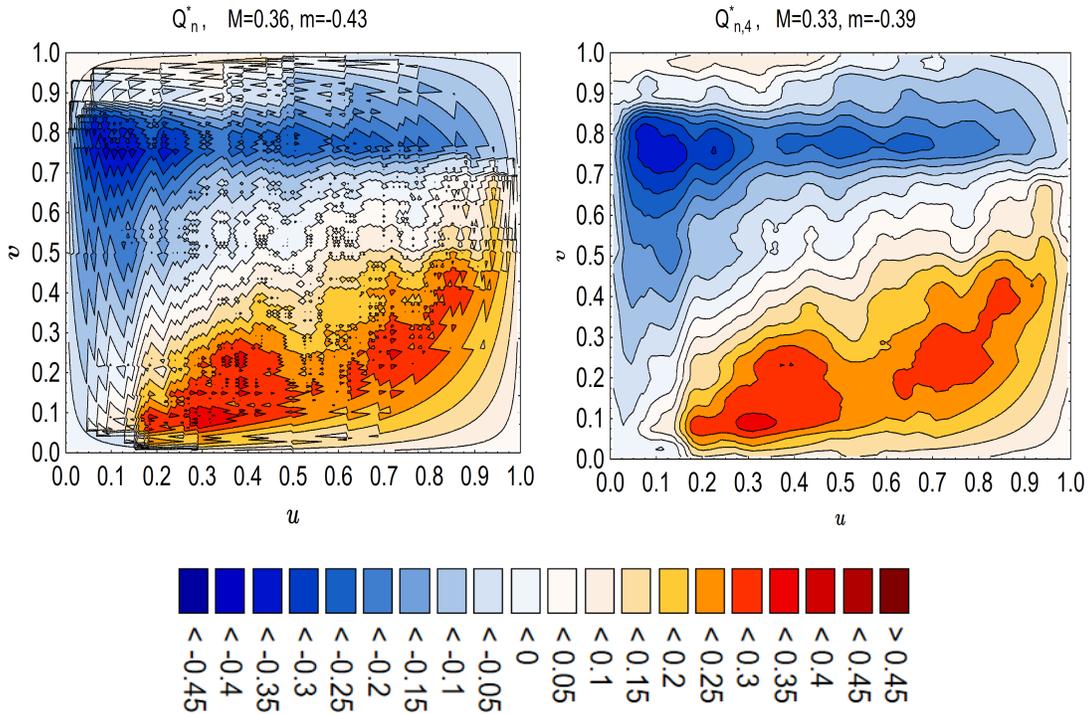} 
\caption{Estimates $Q_n^*$ and $Q_{n,4}^*$ for n=230 aircraft span and speed data.}\label{fig3}
\end{figure}

In Figure \ref{fig3} we show two our estimators $Q_n^*$ and $Q_{n,4}^*$ for these data. The quantile dependence function $q$,  relying on the copula,  separates the dependence structure from any marginal effects. At first glance it can be noticed that the dependence pattern, which has been revealed by the estimated $q$, is much simpler than the above mentioned pictures suggest. Both our displays show two separated regions of strong dependence (positive and negative) of spans and speeds. 
Spans in the range from the second to ninth decile are positively correlated with speeds lying below their median. For speed above the median and below the ninth decile, a strong negative trend is exhibited. There is also a large region of $u$-quantiles and $v$-quantiles in which the variables span and speed seem to be unrelated.

It is visible at first glance that in some regions of points $(u,v)$'s of the right hand panel in Figure 3 the absolute value of $\sqrt n Q^*_{n,4}(u,v)$ exceeds values which, in view of Proposition 2, are highly improbable under the null hypothesis. For instance, $\sqrt n Q^*_{n,4}(0.106,0.742) \approx -5.9.$
Arguing formally, all $p$-values of our new statistics ${\cal D}_{\kappa,s,n}^*,\;s=0,4,\kappa=0.025$, ${\cal L}_{\epsilon,r,n}^*,\;r=2,6, \epsilon=0.01$,  and related ${\cal M}_n^*$ are practically 0.

\section{Conclusion}
\label{S7}

In this article, we proposed a framework for measuring and visualizing the dependence structure, and constructing formal tests of independence of two  random variables. Our approach exploits the quantile dependence function $q$, a recently introduced local dependence measure.  The function $q$ gives a detailed picture of the underlying dependence structure. It provides a means to carefully examine local association structure at different quantile levels. 
The measure distinguishes between negative and positive quadrant dependence and can be immediately generalized to the multivariate case. We have proposed three new tests based on simple and useful nonparametric estimators of the measure. Estimating the measure naturally leads to some weighted copula processes. Our test statistics are based on classical supremum-type and integral-type functionals of the related processes. Both the measure and the corresponding test statistics are invariant under strictly increasing transformations of observations. These statistics are easy and fast to calculate.
A more refined solution has been proposed as well.
Some consistency results are stated and proved, and extensive evidence on stability of empirical powers of new solutions is provided. Finally, we have applied our approach to analyze Saviotti aircraft data. This application shows the usefulness of the proposed inference tools in two respects: as a simple and reliable graphical device allowing for visual inspection of regions of evident departures from independence and formal tests verifying validity of the independence structure.

\bibliographystyle{apalike}
\setlength{\bibsep}{6pt}
\bibliography{Bibliography-VoA}

%\includepdf[pages=-,scale=1,offset=42 0]{Appendices.pdf}

\section{Supplementary Materials}

\noindent
{\large{\bf Appendix A:   Proofs}}\\

\noindent 
 {\bf A.1 Proof of Proposition 2}\\ 

\noindent
The first part of (8) follows immediately from Theorem 3 in Fermanian et al. (2004) and the forms of the limiting process and pertaining covariance. 

To prove the second statement of (8) let us introduce  auxiliary rank process and recall the abbreviated notation for our weight function
$$
\mathbb{C}_n^{(1)}(u,v) =\sqrt n \{C_n(u,v) - uv\},\;\;\;w(u,v)=\frac{1}{\sqrt{uv(1-u)(1-v)}},
\eqno(A.1)
$$
where $C_n(u,v)$ is given by (3). With these notation we have: for each fixed $(u,v) \in (0,1)^2$ it holds that 
$w(u,v)\mathbb{C}_n^{(1)}(u,v)=\sqrt n Q_n(u,v)=\sqrt n w(u,v)N_n^{(1)}(u,v)$. Hence $\sqrt n w(u,v)N_n^{(1)}(u,v) \stackrel{D}{\rightarrow} N(0,1)$, provided that ${\bf H_0}$ is true. 
The last statement is key observation to complete a proof of the second part of (8). Indeed, consider  the succeeding expression $N_n^{(2)}(u,v)$ appearing in $N_n^*(u,v)$, which in turn defines $Q_n^*(u,v)$. We have
$$
N_n^{(2)}(u,v)=  N_n^{(1)}(u,v) -r_n^{(2)},\;\;\mbox{where}\;\;r_n^{(2)}=\frac{1}{n}\lfloor (n+1)u\rfloor -u,
\;\;\mbox{and}\;\;|r_n^{(2)}|\leq \frac{1}{n}\;\;\mbox{for}\;\;u\in [0,1],
\eqno(A.2)
$$
while $\lfloor \bullet \rfloor$ denotes the integer part of the number $\bullet $. Analogously, 
$$
N_n^{(4)}(u,v)=  N_n^{(1)}(u,v) -r_n^{(4)},\;\;\;\mbox{where}\;\;\;r_n^{(4)}=\frac{1}{n}\lfloor (n+1)v\rfloor -v,
\;\;\mbox{and}\;\;|r_n^{(4)}|\leq \frac{1}{n}\;\;\mbox{for}\;\;v\in [0,1],
\eqno(A.3)
$$
and
$$
N_n^{(3)}(u,v)=-N_n^{(1)}(u,v) + N_n^{(2)}(u,v) + N_n^{(4)}(u,v) = N_n^{(1)}(u,v)-r_n^{(2)}-r_n^{(4)}.
\eqno(A.4)
$$
Obviously, the relations (A.2)-(A.4) hold irrespective of ${\bf H_0}$ is true or not.
Under ${\bf H_0}$, asymptotic normality of $\sqrt n w(u,v)N_n^{(1)}(u,v)$ and the above yield the required statement for $\sqrt n Q_n^*(u,v)$. 

Now we shall show that $Q_n^*$ and $Q_{n,s}^*$ are close enough to infer the last statement in (8) from the middle one. For this purpose take any $(u_0,v_0) \in (0,1)^2$. We claim that for $(u_0,v_0)$ it holds that: there exists $n_0 \in \mathbb{N}$ and a constant $c(n_0,u_0,v_0,s)$ such that for all $n \geq n_0$ 
$$
\bigl|Q_n^*(u_0,v_0) - Q_{n,s}^*(u_0,v_0)\bigr| \leq \frac{1}{n} c(n_0,u_0,v_0,s).
\eqno(A.5)
$$
Without loss of generality assume that $(u_0,v_0) \in (0,1/2)^2$. Set
$$
B_{n,s}(u_0,v_0)=\Bigl[u_0-\frac{s}{n+1},u_0+\frac{s}{n+1}\Bigr]\times \Bigl[v_0-\frac{s}{n+1},v_0+\frac{s}{n+1}\Bigr].
$$
Then, there exists $n_0 \in \mathbb{N}$ such that $B_{n,s}(u_0,v_0) \subset (0,1/2)^2.$ Moreover, for any $n \in \mathbb{N}$ it holds that $B_{n+1,s}(u_0,v_0) \subset
B_{n,s}(u_0,v_0)$. Hence, for any $n \geq n_0$
$$
\sup_{(u,v) \in B_{n,s}(u_0,v_0)} \bigl|N_n^*(u,v)-N_n^*(u_0,v_0)\bigr| = \sup_{(u,v) \in B_{n,s}(u_0,v_0)}\bigl|C_n(u,v)- uv - C_n(u_0,v_0) +u_0 v_0\bigr|  
$$
$$
\leq \sup_{(u,v) \in B_{n,s}(u_0,v_0)} \bigl|C_n(u,v) - C_n(u_0,v_0)\bigr| + \sup_{(u,v) \in B_{n,s}(u_0,v_0)} \bigl|uv - u_0 v_0\bigr|
\eqno(A.6)
$$
$$
\leq \frac{s}{n+1} + \frac{su_0}{n+1} + \frac{sv_0}{n+1} + \frac{s^2}{(n+1)^2} \leq \frac{3s}{n+1}.
$$
Since $w(u,v)$ is continuous and bounded on $B_{n_0,s}(u_0,v_0)$, therefore there exists $\bar {c}(n_0,u_0,v_0,s)$ such that for all $n \geq n_0$
$$
\sup_{(u,v) \in B_{n,s}(u_0,v_0)} \bigl|Q_n^*(u,v)-Q_n^*(u_0,v_0)\bigr| \leq \bar {c}(n_0,u_0,v_0,s) \sup_{(u,v) \in B_{n,s}(u_0,v_0)} \bigl|N_n^*(u,v)-N_n^*(u_0,v_0)\bigr|.
\eqno(A.7)
$$
By (A.6) and (A.7),
$$
\bigl|Q_n^*(u_0,v_0) - Q_{n,s}^*(u_0,v_0)\bigr| \leq \int_{B_{n,s}(u_0,v_0)}\bigl(\frac{2s}{n+1}\bigr)^{-2}\bigl|Q_n^*(u,v)-Q_n^*(u_0,v_0)\bigr|dudv
$$
$$
\leq \sup_{(u,v) \in B_{n,s}(u_0,v_0)} \bigl|Q_n^*(u,v)-Q_n^*(u_0,v_0)\bigr| \leq \frac{3s}{n+1} \bar{c} (n_0,u_0,v_0,s).
$$
This proves (A.5) and finally yields the last conclusion in (8). \hfill $\Box$\\

In view of (5) and an analogous relations for $N_n(u,v)$, the zero set of the denominator in (6) is included in the zero set of the numerators in (6). Hence, we can additionally define $Q_n$ and $Q_n^*$ on this set to be 0. With this convention, the sample paths of $Q_n$ and $Q_n^*$ are bounded on $[0,1]^2$. This enables to treat $Q_n$ and $Q_n^*$ as random elements with values in ${\ell}^{\infty}([0,1]^2)$. This observation helps us to use below some ready results on weighted copula process.\\

%\newpage
\noindent
{\bf A.2 Proof of Proposition 3}

\vspace{0.4cm}
\noindent   
We shall consider the integral ${\cal L}_{\epsilon,r,n}^*$ on  four subsets of $A({\epsilon})$ in (10), separately.
On the set $A^{(1)}(\epsilon)=[0,1/2]^2\setminus [0,\epsilon]^2$ it holds that $\sqrt n  Q_n^*(u,v)=w(u,v)\mathbb{C}_n^{(1)}(u,v)$, where $\mathbb{C}_n^{(1)}(u,v)$ is defined in (A.1). On the set $A^{(3)}(\epsilon)=[1/2,1]^2\setminus [1-\epsilon,1]^2$ we have
$$
\sqrt n \Bigl\{\int_{A^{(3)}(\epsilon)}|Q_n^*(u,v)|^r dudv\Bigr\}^{1/r} = \sqrt n \Bigl\{\int_{A^{(3)}(\epsilon)}|w(u,v)N_n^{(3)}(u,v)|^r dudv\Bigr\}^{1/r}=
$$
$$
\sqrt n \Bigl\{\int_{A^{(3)}(\epsilon)}\Bigl[w(u,v)\Bigl|\frac{1}{n}\sum_{i=1}^n{\bf 1}\Bigl(\frac{R_i}{n+1}>u,\frac{S_i}{n+1}>v\Bigr)-(1-u)(1-v)\Bigr|\Bigr]^r dudv\Bigr\}^{1/r}=
$$
$$
\sqrt n \Bigl\{\int_{A^{(1)}(\epsilon)}\Bigl[w(u,v)\Bigl|\frac{1}{n}\sum_{i=1}^n{\bf 1}\Bigl(\frac{R_i}{n+1}>1-s,\frac{S_i}{n+1}>1-t\Bigr)-st\Bigr|\Bigr]^r dsdt\Bigr\}^{1/r}.
$$
Moreover, the process $\sqrt n \Bigl\{\frac{1}{n}\sum_{i=1}^n{\bf 1}\Bigl(\frac{R_i}{n+1}>1-s,\frac{S_i}{n+1}>1-t\Bigr)-st\Bigr\}$ has the same distribution as the process 
$\mathbb{C}_n^{(3)}(s-,t-)$, where $\mathbb{C}_n^{(3)}$ is the rank process for the sample $(f(X_i),g(Y_i)),\;i=1,...,n$, where the functions $f$ and $g$ are strictly decreasing. It is so because the ranks of the transformed observations $R_i'$ and $S_i'$, say, are related to $R_i$ and $S_i$ as follows $R_i'=n+1-R_i,\;S_i'=n+1-S_i$. For the two remaining subsets of the integration area $A({\epsilon})$ similar argument applies. It shows that asymptotic behavior of ${\cal L}_{\epsilon,r,n}^*$ is determined via pertaining asymptotics of the variables 
$$
\Bigl\{\int_{A^{(1)}(\epsilon)}\Bigl|w(u,v)\mathbb{C}_n^{(j)}(u,v)\Bigr|^rdudv\Bigr\}^{1/r},\;\;j=1,...,4.
$$
Moreover, 
$$
\sup_{(u,v)\in [0,1]^2\setminus J_n} Q_n^*(u,v)=O\bigl(\frac{1}{n}\bigr)\;\;\;\mbox{where}\;\;\;J_n=\bigl[\frac{1}{n+1},1-\frac{1}{n+1}\bigr]^2.
$$
On the other hand, under the  Assumption 0, by Theorem 2.2 of Berghaus et al. (2017), the empirical copula process 
$$
\hat {\mathbb{C}}_n(u,v)=\sqrt n  \{C_n(u,v)-C(u,v)\}
\eqno(A.8)
$$ 
for $ (u,v) \in J_n$ is well approximated, in some weighted supremum norm, by pertaining  bivariate empirical process $\bar {\mathbb{C}}_n(u,v)$. To define $\bar {\mathbb{C}}_n(u,v)$ set $U_i=F(X_i), V_i=G(Y_i), i=1,...,n$, and
$$
D_n(u,v)=\frac{1}{n} \sum_{i=1}^n {\bf 1}(U_i \leq u,V_i \leq v).
\eqno(A.9)
$$
Next, introduce the (unobservable) empirical process $\alpha_n$, based on $(U_1,V_1),...,(U_n,V_n)$, 
$$
\alpha_n(u,v)=\sqrt n \{D_n(u,v)-C(u,v)\},
\eqno(A.10)
$$
and finally define
$$
\bar {\mathbb{C}}_n(u,v) = \alpha_n(u,v)-\dot{C}_1(u,v)\alpha_n(u,1)-\dot{C}_2(u,v)\alpha_n(1,v),
\eqno(A.11)
$$
where $\dot{C}_1(u,v)=\frac{\partial}{\partial u}C(u,v)$ and $\dot{C}_2(u,v)=\frac{\partial}{\partial v}C(u,v)$.
With this notation, %it holds that
$$
\sup_{(u,v)\in J_n}\Bigl|\frac{ \hat {\mathbb{C}}_n(u,v)}{g_{\omega}(u,v)} -  \frac{ \bar {\mathbb{C}}_n(u,v)}{g_{\omega}(u,v)}\Bigr|=o_P(1),
$$
where $g_{\omega}(u,v)= [\min\{u,v,1-u,1-v\}]^{\omega}$ and $\omega \in (0,1/2)$. Moreover, it holds that ${\bar {\mathbb{C}}_n(u,v)}/{g_{\omega}(u,v)}$ converges weakly in $\ell^{\infty}([0,1]^2)$, equipped with the supremum norm, to centered Gaussian process.

The above implies that asymptotic behavior of the linear functional ${\cal L}_{\epsilon,r,n}^*$
is determined by respective asymptotics of the weighted bivariate empirical process and its pertaining variants, provided that 
we are well controlling the magnitude of $I_r(\epsilon)=\int_{A({\epsilon})}[w(u,v)g_{\omega}(u,v)]^r dudv$. Note also that asymptotic behavior of 
$$
\Bigl\{\int_{A(\epsilon)}\bigl|w(u,v)\hat{\mathbb C}_n(u,v) + \sqrt n w(u,v)[C(u,v)-uv]\bigr|^r\Bigr\}^{1/r}.
$$
is decisive to the asymptotic power of ${\cal L}_{\epsilon,r,n}^*$ under the alternative pertaining to $C$.
By symmetries of $w$ and $q_{\omega}$, we have
$$
I_r(\epsilon)=\int_{A(\epsilon)}[w(u,v)g_{\omega}(u,v)]^r dudv= 8 \int_{\epsilon}^{1/2} \int_0^v \Bigl[\frac{u^{\omega}}{\sqrt{uv(1-u)(1-v)}}\Bigr]^r dudv.
$$
In the last integral the behavior of $\int_{\epsilon}^{1/2} \int_0 ^v [u^{\omega}/\sqrt{uv}]^r dudv$ is crucial. Hence, the requirement $\omega > 1/2 -1/r$ 
follows. This implies our assumption $r>2.$

The above yields
$$
I_r(\epsilon_n) \leq c_1 \int_{\epsilon_n}^{1/2} \frac{v^{\omega r - r +1}}{\omega r -r/2 +1}dv \leq c_2 {\epsilon_n}^{{\omega}r -r} \leq c_2 {\epsilon_n}^{{-r/2}-1},
$$
where $c_1$ and $c_2$ are absolute constants. Hence, under our assumptions on ${\epsilon_n}$, $I_r(\epsilon_n)=o(\sqrt n)$. The above implies that asymptotics of 
${\cal L}_{r,n}^*(\epsilon_n)$, under $C$, is determined by the two terms: a random component, which is at most $o_P(\sqrt n)$ and an asymptotic shift, which is at least $O(\sqrt n)$. This ensures the consistency. \hfill $\Box$\\
\newpage
\noindent
{\bf A.3 Proof of Proposition 4}

\vspace{0.4cm}
\noindent
Due to 1. of the Assumption 0, by Proposition 3.1 of Segers (2012), the process $\hat{\mathbb{C}}_n$, given in the equation (A.8), converges weakly in $\ell^{\infty}([0,1]^2)$ to the Gaussian process $\mathbb{C}$, defined by (3.1) in Segers (2012). This implies that for any $\kappa=\kappa_n$, $\kappa_n \to 0$ as $n \to \infty$, 
$$
\sup_{(u,v) \in [\kappa,1-\kappa]^2}  w(u,v)|\hat{\mathbb{C}}_n(u,v)|=O_P(1/{\kappa}_n).
\eqno(A.12)
$$
On the other hand, by (6), 
$$
\sqrt n Q_n(u,v)= w(u,v)[\hat{\mathbb{C}}_n(u,v)] + \sqrt n w(u,v)[C(u,v)-uv].
\eqno(A.13)
$$
Under the alternative $C$, the second term in (A.13) is, in $[\kappa_n,1-\kappa_n]^2$, at least $O({\sqrt n})$. Therefore,  the assertion (A.12) along with the assumptions on $\kappa_n$ yield the consistency of $ \sup_{(u,v) \in [\kappa,1-\kappa]^2}|\sqrt n Q_n(u,v)|$. 

The case of $\sqrt n Q^*_n(u,v)$ can be treated similarly, as by (A.2)-(A.4), in an analogue of (A.13) an immaterial extra term, being at most 
$O(1/(\sqrt{n}\kappa_n))$, appears, only. Hence, for $\kappa_n$'s under consideration,  the consistency of ${\cal D}^*_{\kappa,0,n}$ follows.

To prove consistency of the test rejecting ${\bf H_0}$ for large values of ${\cal D}^*_{\kappa,s,n}$ observe that always it holds that 
${\cal D}^*_{\kappa,s,n} \leq {\cal D}^*_{\kappa,0,n}$. Let $c_{\alpha,n}$ be a critical value of $\alpha$-level test based on ${\cal D}^*_{\kappa,0,n}$. By the above, $c_{\alpha,n}=O(1/\kappa_n)$. On the other hand, by (A.5), under any alternative defined by the underlying $C$, 
$$
\lim_{n \to \infty} P\bigl({\cal D}^*_{\kappa,s,n} \geq c_{\alpha,n}\bigr)=\lim_{n \to \infty} P\bigl({\cal D}^*_{\kappa,0,n} \geq c_{\alpha,n}+O(\frac{1}{\sqrt n})\bigr)=1.
$$ 
In view of  ${\cal D}^*_{\kappa,s,n} \leq {\cal D}^*_{\kappa,0,n}$ the proof of the consistency of the smoothed variant  is concluded.
\hfill $\Box$\\

\vspace{1.5cm}

\noindent
{\large {\bf Appendix B:   Data set referenced in the article}}\\

\noindent
Aircraft span and speed data, from the third period, 1956-1984, are available in electronic form from
A.W. Bowman and A. Azzalini  (2007). {\it R package ‘sm’: Nonparametric smoothing methods (version 2.2)};   https://cran.r-project.org/web/packages/sm/sm.pdf\\
\\
\\
\noindent
{\large {\bf Appendix C: Description and Codes}}\\

\noindent
In this Section we provide some $C$ codes for computation of ${\cal Q}_n^*$, ${\cal Q}_{n,s}^*$, ${\cal L}^*_{\epsilon,r,n},\;{\cal D}^*_{\kappa,0,n},\;{\cal D}^*_{\kappa,s,n}$, and comment on calculation of ${\cal M}_n^*$. We start with a preliminary information.\\
\\
\noindent
{\bf C.1. Preliminaries}\\

\noindent
{\bf Calculation of $Q^*_n$}\\

\noindent
Let us denote by $R=(R_1,...,R_n)$ and $S=(S_1,...,S_n)$ the  vectors of ranks related to $(X_1,...,X_n)$ and $(Y_1,...,Y_n)$, respectively.
Set $R'_i=n+1-R_i$ and $S'_i=n+1-S_i$, $i=1,...,n$. Next, introduce the notation 
$R'=(R'_1,...,R'_n)$ and $S'=(S'_1,...,S'_n)$ for vectors of transformed ranks. Additionally define $u'=1-u$ and $v'=1-v$. Recall also that, using $R$ and $S$, we consider empirical copula of the form 
$$ 
C_n(u,v)= C_n(u,v;R,S)=\frac{1}{n}\sum_{i=1}^n {\bf 1}\Bigl(\frac{R_i}{n+1} \leq u,\frac{S_i}{n+1} \leq v\Bigr).
\eqno(C.1)
$$
With these notations, the functions $N_n^{(k)}(u,v),\;k=1,...,4$, appearing in the definition of $N_n^*(u,v)$, and hence in the formula (6) for $Q_n^*(u,v)$, have the following alternative forms
\\
$$
N^{(1)}_n(u,v)= N^{(1)}_n(u,v;R,S)= \frac{1}{n}\sum_{i=1}^n {\bf 1}\bigl(\frac{R_i}{n+1} \leq u, \frac{S_i}{n+1} \leq v\bigr) - uv=C_n(u,v;R,S)-uv, 
$$
$$
N^{(2)}_n(u,v)=uv' -\frac{1}{n}\sum_{i=1}^n {\bf 1}\bigl(\frac{R_i}{n+1} \leq u,\frac{S'_i}{n+1}<v'\bigr)=-N_n^{(1)}(u,v';R,S'),
\eqno(C.2)
$$
$$
N^{(3)}_n(u,v)= \frac{1}{n}\sum_{i=1}^n {\bf 1}\bigl(\frac{R'_i}{n+1} <u', \frac{S'_i}{n+1} < v'\bigr) - u'v'=N^{(1)}_n(u',v';R',S'), 
$$
$$
N^{(4)}_n(u,v)= u'v -\frac{1}{n}\sum_{i=1}^n {\bf 1}\bigl(\frac{R'_i}{n+1} <u',\frac{S_i}{n+1}\leq v\bigr)=-N^{(1)}_n(u',v;R',S),
$$ 
for any $(u,v)\notin\{1/(n+1),...,n/(n+1)\}^2$. That is why we calculate the transformed ranks in our computer program and we calculate empirical copula for ranks and transformed ranks. Tables "Ctab", "Ctabs12", "Ctabs22", "Ctabs21" contain values of the empirical copulas $C_n(u,v;R,S)$, $C_n(u,v';R,S')$, $C_n(u',v';R',S')$, $C_n(u',v;R',S)$,  corresponding to a grid of points $(u_i,v_j)$'s, where 
$$
u_i=\frac{i+0.5}{n+1}\;\;\  \mbox{and}\;\;\  v_j=\frac{j+0.5}{n+1} \;\;\;\mbox{for}\;\;i=0,...,n,\;j=0,...,n.
$$
Notice that a calculation of  $C_n(u,v)$, on the  grid $(u_i,v_j)$, $i=0,...,n$, $j=0,...,n$, using the initial formula (C.1)
is completely ineffective since for every point from the  grid we have to calculate a sum of $n$ indicators. The computational complexity of that approach is $n^3$. Much better method is the following recursion.
If we sort vectors $(R_1,S_1),...,(R_n,S_n)$ in ascending order according to the first coordinate, we obtain vectors $(1,S_{[1]}),...,(1,S_{[n]})$, where $S_{[k]}$ is the rank in the vector $S$ corresponding to the rank $k$ in the vector $R$. 
Observe that  for all $j=0,1,...,n$ we have 
$C_n(u_0,v_j;R,S)=0$. Moreover,  for all $i=1,...,n$, and  for $j=0,...,n$ it holds that 
\begin{equation}\nonumber
C_n(u_i,v_j;R,S)=\left\{ \begin{array}{l}
C_n(u_{i-1},v_j;R,S) \qquad \mbox{if} \;\;\;j<S_{[i]},\\
\\
C_n(
u_{i-1},v_j;R,S)+1/n\qquad \mbox{if} \;\;\;j\geqslant S_{[i]}.\\
\end{array}\right.
\end{equation}
Let us explain this recursion using the following example, in which we took $n=10$. The resulting $S_{[k]}$'s were as follows: 3, 6, 2, 9, 4, 1, 7, 5, 8, 10. 
The entries of the presented $11 \times 11$ table  are equal to respective values of $C_n(u_i,v_j;R,S)$ for $i=0,...,n$, $j=0,...,n$. It is useful to imagine that succeeding vertical and horizontal lines of the table are located at points $l/(n+1),\;l=0,...,n+1.$

\begin{figure}[!h]
\centering
\includegraphics[trim = 0mm 120mm 0mm 0mm, scale=0.7]{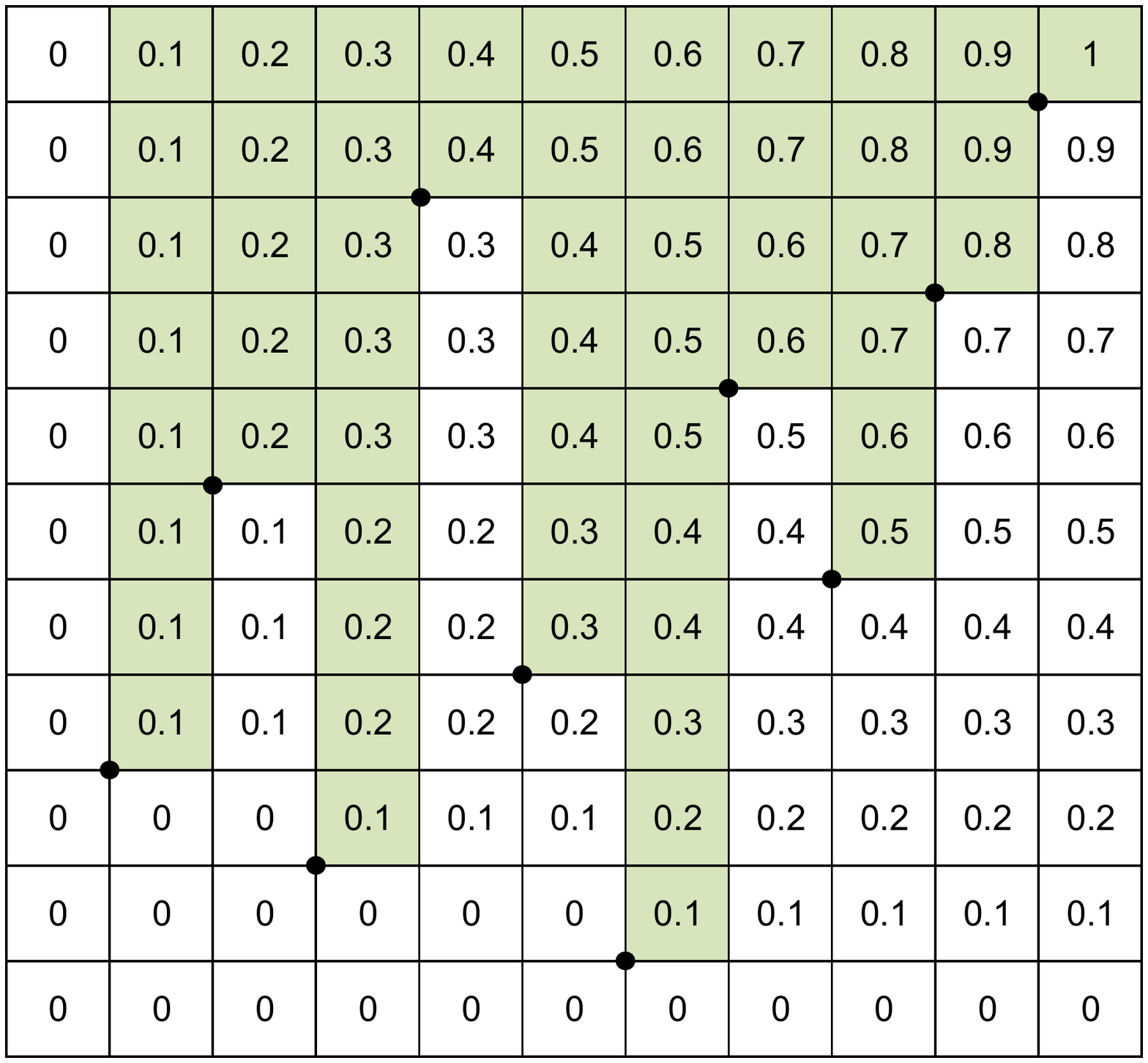} 
\end{figure}

\noindent  
All entries in the first column are equal to $0$. In next columns the values on white background are equal to the first value on the left and the values on green background are equal to the first value on the left plus $1/n$. When looking from the left to the right,  black dots in the table are the sorted pseudo-observations $(k/(n+1),S_{[k]}/(n+1)),\;k=1,...,n$. The computational complexity of that approach is $n^2$. 
Tables "T", "Ts12", "Ts22", "Ts21" contain second coordinates of the sorted vectors $((R_1,S_1),...,(R_n,S_n))$, $((R_1,S'_1),...,(R_n,S'_n))$, $((R'_1,S'_1),...,(R'_n,S'_n))$, $((R'_1,S_1),...,(R'_n,S_n))$, according to the first coordinate, respectively. Using the above recursion, we use them to calculate $C_n(u,v;R,S)$, $C_n(u,v';R,S')$, $C_n(u',v';R',S')$, $C_n(u',v;R',S)$ on the grid $(u_i,v_j)$, $i=0,...,n$, $j=0,...,n$. In view of (C.2), this allows to  calculate $Q^*_n(u,v)$ on the grid. Pertaining values are collected by our program in a table "KS".
\\

\newpage
\noindent
{\bf Calculation of $Q^*_{n}$ and  $Q^*_{n,s}$}

\vspace{0.4cm}
\noindent
We calculate 
\begin{equation}\nonumber
Q_{n,s}^*(u,v) = \Bigl(\frac {2s}{n+1}\Bigr)^{-2} \int_{u-\frac{s}{n+1}}^{u+\frac{s}{n+1}} \int_{v-\frac{s}{n+1}}^{v+\frac{s}{n+1}} Q_n^*(x,y)dy dx
\end{equation}
on the grid $(u_i,v_j)$, $i=0,...,n$, $j=0,...,n$ numerically, i.e. we apply the approximation
\begin{equation}\nonumber
Q_{n,s}^*(u_i,v_j) \approx \frac{1}{(2s+1)^2}\sum\limits_{(k,l)\in\{1,...,n\}^2:|k-i|\leqslant s, |l-j|\leqslant s} Q_{n}^*(u_k,v_l).
\end{equation}
The values $Q_{n,s}^*(u_i,v_j)$ for $j=0,...,n$, $j=0,...,n$ are collected in a table "KWyg". \\

Notice also that $Q_{n}^*(u_i,v_j)=Q_{n,0}^*(u_i,v_j)$. So, in this way, both estimators of the quantile dependence function $q$ can be calculated. \\ 
\\
\\
\noindent
{\bf Calculation of ${\cal L}^*_{\epsilon,r,n}$}\\

\noindent
Using the definition of $A(\epsilon)$, given by the formula (10) of the paper, set
$$
K_{\epsilon,n}=\{(i,j)\in \{0,...,n\}^2: (u_i,v_j)\in A(\epsilon)\}.
$$
We calculate ${\cal L}^*_{\epsilon,r,n}$ numerically,  i.e. we approximate
$$
{\cal L}^*_{\epsilon,r,n} \approx \frac{\sqrt n}{(n+1)^{2}}\left(\sum\limits_{(i,j) \in K_{\epsilon,n}} \big|Q_n^*(u_i,v_j)\big|^r\right)^{1/r}.
$$
In the computer code the approximate value of ${\cal L}^*_{\epsilon,r,n}$ is denoted by "I".\\
\\
\\
\noindent
{\bf Calculation of } ${\cal D}^*_{\kappa,0,n}$ {\bf and} ${\cal D}^*_{\kappa,s,n}$
\\

\noindent
Let us denote 
$$
J_{\kappa,n}=\{(i,j)\in \{0,...,n\}^2: (u_i,v_j)\in [\kappa,1-\kappa]^2\}.
$$
We calculate ${\cal D}^*_{\kappa,0,n}$ and ${\cal D}^*_{\kappa,s,n}$ numerically, i.e.
$${\cal D}^*_{\kappa,0,n} \approx \max\limits_{(i,j) \in J_{\kappa,n}} |Q_{n}^*(u_i,v_j)|, \ \ \ \ \ \ \    {\cal D}^*_{\kappa,s,n} \approx \max\limits_{(i,j) \in J_{\kappa,n}} |Q_{n,s}^*(u_i,v_j)|. $$
In the computer code the approximate values of ${\cal D}^*_{\kappa,0,n}$ and ${\cal D}^*_{\kappa,s,n}$ are denoted by "D0" and "Ds", respectively.\\
\\
\\
\noindent
{\bf Calculation of  ${\cal M}^*_{n}$}\\

\noindent
A scheme of our program in this case was as follows. Given observations $(X_1,Y_1),...,(X_n,Y_n)$, we calculated pertaining values of statistics,  L = ${\cal L}^*_{\epsilon,r,n}$ and H = HHG, say, where HHG denotes rank based variant of test introduced in Heller et al. (2013). Note that basic part of a C code for the statistic HHG is given in Section 2 of Supplementary Material for this paper.

To estimate $p$-values of both ingredients of ${\cal M}^*_{n}$ we have applied Monte Carlo method. Namely, we have generated MC = 100 000 auxiliary samples of size $n$ from uniform distribution on $(0,1)^2$ and calculated for them respective values of both statistics. Next, we sorted $MC$ values of ${\cal L}^*_{\epsilon,r,n}$ in ascending order obtaining $L_{(1)},....,L_{(MC)}$. Similarly, we obtained sorted values $H_{(1)},....,H_{(MC)}$ of HHG. Empirical $p$-values for the obtained values $L$ and $H$ were estimated as follows
$$
p_1=\frac{1}{MC}\sum\limits_{i=1}^{MC} {\bf 1}(L<L_{(i)})\;\;\;\mbox{and}\;\;\;p_2=\frac{1}{MC}\sum\limits_{i=1}^{MC} {\bf 1}(H<H_{(i)}).
$$
Finally, we calculated ${\cal M}^*_{n}=\min\{p_1,p_2\}$.

%\end{section}

\noindent
{{\large \bf C.2. Codes}}\\

\noindent
Below we give $C$ codes for computation of ${\cal Q}_n^*$, ${\cal Q}_{n,s}^*$, ${\cal L}^*_{\epsilon,r,n},\;{\cal D}^*_{\kappa,0,n},\;{\cal D}^*_{\kappa,s,n}$. \\


\section*{Supplementary material}

```csharp
// We use MathNet.Numerics.dll and Science.dll
using System;
using System.Collections.Generic;
using System.Linq;
using System.Text;
using System.IO;
using Science.Mathematics.Calculus;
using MathNet.Numerics.Random;
using MathNet.Numerics.Distributions;
using MathNet.Numerics.LinearAlgebra;

// reading from file the ranks of X and Y (function "ReadFromFile" defined at the end of this code)
static double[] Rx = ReadFromFile(X_Ranks);
static double[] Ry = ReadFromFile(Y_Ranks);

static int n = Rx.Length; // calculation of sample size

static void Main(string[] args)
{
    double r = 6.0;      // value $r$ for the statistics $L^*_{\varepsilon,r,n}$
    double epsilon = 0.01; // value $\varepsilon$ for the statistics $L^*_{\varepsilon,r,n}$
    double kappa = 0.025; // value $\kappa$ for the statistics $D^*_{\kappa,0,n}$ and $D^*_{\kappa,s,n}$
    int s = 4;    // value $s$ for the statistics $D^*_{\kappa,s,n}$

    // auxiliary tables for calculations
    double[,] Ctab = new double[n + 1, n + 1];
    double[,] Ctabs22 = new double[n + 1, n + 1];
    double[,] Ctabs12 = new double[n + 1, n + 1];
    double[,] Ctabs21 = new double[n + 1, n + 1];
    double[] Rsx = new double[n];
    double[] Rsy = new double[n];
    double[] T = new double[n];
    double[] Ts22 = new double[n];
    double[] Ts12 = new double[n];
    double[] Ts21 = new double[n];
    double[,] KS = new double[n + 1, n + 1];
    double[,] Kwyg = new double[n, n];

    // filling the tables with zeros (function "zero" defined at the end of this code)
    zero(Rsy);
    zero(Rsx);
    zero(T);
    zero(Ts22);
    zero(Ts12);
    zero(Ts21);
    zero(KS);
    zero(Kwyg);
    zero(Ctab);
    zero(Ctabs22);
    zero(Ctabs12);
    zero(Ctabs21);

    // calculation of transformed ranks
    for (int i = 0; i < n; i++)
    {
        Rsx[i] = (double)(n + 1) - Rx[i];
    }

    for (int i = 0; i < n; i++)
    {
        Rsy[i] = (double)(n + 1) - Ry[i];
    }
```

```csharp
// filling auxiliary tables with ranks and transformed ranks
for (int i = 0; i < n; i++)
{
    T[(int)Rx[i] - 1] = Ry[i];
}

for (int i = 0; i < n; i++)
{
    Ts22[(int)Rsx[i] - 1] = Rsy[i];
}

for (int i = 0; i < n; i++)
{
    Ts12[(int)Rx[i] - 1] = Rsy[i];
}

for (int i = 0; i < n; i++)
{
    Ts21[(int)Rsx[i] - 1] = Ry[i];
}


// calculation of empirical copula on the grid, for ranks and transformed ranks
for (int i = 1; i <= (double)(n + 1) / 2.0; i++)
{
    for (int j = 0; j < T[i - 1] && j <= (double)(n + 1) / 2.0; j++)
    {
        Ctab[i, j] = Ctab[i - 1, j];
    }

    for (int j = (int)T[i - 1]; j <= (double)(n + 1) / 2.0; j++)
    {
        Ctab[i, j] = Ctab[i - 1, j] + 1.0 / (double)n;
    }
}

for (int i = 1; i <= (double)(n + 1) / 2.0; i++)
{
    for (int j = 0; j < Ts22[i - 1] && j <= (double)(n + 1) / 2.0; j++)
    {
        Ctabs22[i, j] = Ctabs22[i - 1, j];
    }

    for (int j = (int)Ts22[i - 1]; j <= (double)(n + 1) / 2.0; j++)
    {
        Ctabs22[i, j] = Ctabs22[i - 1, j] + 1.0 / (double)n;
    }
}

for (int i = 1; i <= (double)(n + 1) / 2.0; i++)
{
    for (int j = 0; j < Ts12[i - 1] && j <= (double)(n + 1) / 2.0; j++)
    {
        Ctabs12[i, j] = Ctabs12[i - 1, j];
    }

    for (int j = (int)Ts12[i - 1]; j <= (double)(n + 1) / 2.0; j++)
    {
        Ctabs12[i, j] = Ctabs12[i - 1, j] + 1.0 / (double)n;
    }
}
```

```csharp
for (int i = 1; i <= (double)(n + 1) / 2.0; i++)
{
    for (int j = 0; j < Ts21[i - 1] && j <= (double)(n + 1) / 2.0; j++)
    {
        Ctabs21[i, j] = Ctabs21[i - 1, j];
    }

    for (int j = (int)Ts21[i - 1]; j <= (double)(n + 1) / 2.0; j++)
    {
        Ctabs21[i, j] = Ctabs21[i - 1, j] + 1.0 / (double)n;
    }
}


// calculation of Q_n^* (function "Wn" defined at the end of this code)
for (int i = 0; i <= (double)(n + 1) / 2.0; i++)
{
    for (int j = 0; j <= (double)(n + 1) / 2.0; j++)
    {
        KS[i, j] = Wn(n, Ctab[i, j], ((double)(i) + 0.5) / (double)(n + 1),
                    ((double)(j) + 0.5) / (double)(n + 1))
    }
}

for (int i = 0; i <= (double)(n + 1) / 2.0; i++)
{
    for (int j = 0; j <= (double)(n + 1) / 2.0; j++)
    {
        KS[n - i, n - j] = Wn(n, Ctabs22[i, j], ((double)(i) + 0.5) / (double)(n + 1),
                            ((double)(j) + 0.5) / (double)(n + 1));
    }
}

for (int i = 0; i <= (double)(n + 1) / 2.0; i++)
{
    for (int j = 0; j <= (double)(n + 1) / 2.0; j++)
    {
        KS[i, n - j] = -Wn(n, Ctabs12[i, j], ((double)(i) + 0.5) / (double)(n + 1),
                        ((double)(j) + 0.5) / (double)(n + 1));
    }
}

for (int i = 0; i <= (double)(n + 1) / 2.0; i++)
{
    for (int j = 0; j <= (double)(n + 1) / 2.0; j++)
    {
        KS[n - i, j] = -Wn(n, Ctabs21[i, j], ((double)(i) + 0.5) / (double)(n + 1),
                        ((double)(j) + 0.5) / (double)(n + 1));
    }
}
```

```
// calculation of $Q^*_{n,s}$
for (int i = 0; i < n; i++)
{
    for (int j = 0; j < n; j++)
    {
        int ll = 0;
        for (int a = i - s; a <= i + s; a++)
        {
            for (int b = j - s; b <= j + s; b++)
            {
                if (a >= 0 && a <= n && b >= 0 && b <= n)
                {
                    Kwyg[i, j] = Kwyg[i, j] + KS[a, b];
                    ll = ll + 1;
                }
            }
        }
        Kwyg[i, j] = Kwyg[i, j] / (double)ll;
    }
}

// calculation of $L^*_{\varepsilon,r,n}$
double I = 0.0;
for (int i = 1; i < n; i++)
{
    for (int j = 1; j < n; j++)
    {
        I = I + Math.Pow(Math.Abs(KS[i, j]), r);
    }
}

for (int i = 1; i < (n+1)*epsilon-0.5; i++)
{
    for (int j = 1; j < (n+1)*epsilon-0.5; j++)
    {
        I = I - Math.Pow(Math.Abs(KS[i, j]), r);
    }
}

for (int i = 1; I < (n+1)*epsilon-0.5; i++)
{
    for (int j =(int)((n+1)*(1.0-epsilon)); j < n; j++)
    {
        I = I - Math.Pow(Math.Abs(KS[i, j]), r);
    }
}

for (int i = (int)((n+1)*(1.0-epsilon)); i < n; i++)
{
    for (int j=1; j < (n+1)*epsilon-0.5; j++)
    {
        I = I - Math.Pow(Math.Abs(KS[i, j]), r);
    }
}

for (int i = (int)((n+1)*(1.0-epsilon)); i < n; i++)
{
    for (int j =(int)((n+1)*(1.0-epsilon)); j < n; j++)
    {
        I = I - Math.Pow(Math.Abs(KS[i, j]), r);
    }
}

I = Math.Pow(I / Math.Pow((double)(n + 1), 2.0), 1.0 / r);
```

```csharp
            // calculation of $D^*_{\kappa,0,n}$
            Double D0 = (double)(-n * n);
            for (int i = 0; i < n; i++)
            {
                for (int j = 0; j < n; j++)
                {
                    if (((double)i + 0.5) / (double)(n + 1) > kappa && ((double)i + 0.5) / (double)(n + 1) <
                        1.0 - kappa && ((double)j + 0.5) / (double)(n + 1) > kappa && ((double)j + 0.5) /
                        (double)(n + 1) < 1.0 - kappa)
                    {
                        double TT = Math.Abs(KS[i, j]);
                        if (D0 < TT)
                            D0 = TT;
                    }
                }
            }

            // calculation of $D^*_{\kappa,s,n}$
            Double Ds = (double)(-n * n);
            for (int i = 0; i < n; i++)
            {
                for (int j = 0; j < n; j++)
                {
                    if (((double)i + 0.5) / (double)(n + 1) > kappa && ((double)i + 0.5) / (double)(n + 1) <
                        1.0 - kappa && ((double)j + 0.5) / (double)(n + 1) > kappa && ((double)j + 0.5) /
                        (double)(n + 1) < 1.0 - kappa)
                    {
                        double TT = Math.Abs(Kwyg[i, j]);
                        if (Ds < TT)
                            Ds = TT;
                    }
                }
            }


            // writing $L^*_{\varepsilon,r,n}$, $D^*_{\kappa,0,n}$ and $D^*_{\kappa,s,n}$ to a file
            System.IO.StreamWriter F3out = new System.IO.StreamWriter(sBase + "Results.txt");

                F3out.Write(I.ToString().Replace(",", ".") + ";");
                F3out.Write(D0.ToString().Replace(",", ".") + ";");
                F3out.Write(Ds.ToString().Replace(",", "."));


    }
```

```csharp
// function "Wn"
 static double Wn(int n, double C, double u, double v)
 {
    double W = 0.0;
    W = Math.Pow(n, 0.5) * (C - u * v) / Math.Pow(u * v * (1.0 - u) * (1.0 - v), 0.5);
    return W;
 }

// function "zero"
 static void zero(double[] t)
 {
    int size = t.Length;
    for (int i = 0; i < size; i++)
        t[i] = 0;
 }

 static void zero(double[,] t)
 {
    int sizeX = t.GetUpperBound(0);
    int sizeY = t.GetUpperBound(1);
    for (int i = 0; i < sizeX + 1; i++)
        for (int j = 0; j < sizeY + 1; j++)
            t[i, j] = 0;
 }

// function "ReadFromFile"
 static double[] ReadFromFile(string sFun)
 {
    System.IO.StreamReader fFun = new System.IO.StreamReader(sBase + sFun);
    string[] stFun = fFun.ReadToEnd().Split('\r');
    double[] dFun = new double[stFun.Length];

    for (int i = 0; i < stFun.Length; i++)
        dFun[i] = Convert.ToDouble(stFun[i]);

    stFun = null;
    fFun.Close();
    System.GC.Collect();

    return dFun;
 }
```


\end{document}